\documentclass[fleqn,a4paper,authoryear]{elsarticle}

\usepackage[a4paper,margin=1in]{geometry}
\usepackage{fancyhdr}
\usepackage{dsfont}
\usepackage[utf8]{inputenc}
\usepackage[fleqn]{amsmath}
\usepackage{amssymb}
\usepackage{booktabs}
\usepackage{longtable}
\usepackage{xcolor}
\usepackage{appendix}
\usepackage{stmaryrd}
\usepackage{subfig}
\usepackage{hyperref}

\usepackage{mathtools}
\usepackage{amssymb}
\usepackage{xcolor}
\usepackage{appendix}
\usepackage{graphicx}
\graphicspath{ {/home/alexandrosstathas/Documents/PhD_topics/PAPER_II/PAPER_II_Submission_docs/image_files_for_review/} }
\usepackage{float}
\usepackage{layout} 

\usepackage{xr-hyper}
\title{Fault friction under thermal pressurization during large coseismic-slip Part I: Numerical analyses}

\journal{-}

\begin{document}

\author[1]{Alexandros Stathas}
\author[1]{Ioannis Stefanou\corref{cor1}}
\ead{ioannis.stefanou@ec-nantes.fr}
\cortext[cor1]{Corresponding author}
\address[1]{Institut de Recherche en Génie Civil et Mécanique (UMR CNRS 6183), Ecole Centrale de Nantes, Nantes, France}

\begin{abstract}
\noindent In this paper, we study the role of thermal pressurization in the frictional response of a fault under large coseismic slip. We investigate the role of the seismic slip velocity, mixture compressibility, characteristic grain size and viscosity parameter in the frictional response of the coupled thermo-hydro-mechanical problem, taking into account the fault's microstructure. Starting from the mass, energy and momentum balance for Cosserat continua we derive the equations of our model. We complete the mathematical description using perfect plasticity and Perzyna viscoplasticity in the material constitutive behavior. We investigate both the rate independent as well as the rate dependent frictional response and compare with existing models found in literature, namely the rate and state friction law (\cite{dieterich1992earthquake},\cite{ruina1983slip}). We show that our model is capable of predicting strain rate hardening and velocity softening without the assumption of a state variable. We observe traveling instabilities inside the layer that lead to oscillations in the fault's frictional response, like in the case of Portevin Le Chatelier (PLC) effect. This behavior is not captured by existing numerical analyses presented in \cite{Rattez2018b,rattez2018numerical,Rattez2018a} and go beyond the established models of uniform shear (\cite{lachenbruch1980frictional}) and shear on a mathematical plane (\cite{rice2006heating}), which predict a strictly monotonous behavior during shearing. Recent experimental analyses, which have managed to insulate thermal pressurization from other weakening mechanisms (\cite{Badt2020}), corroborate our numerical results.
\end{abstract}

\maketitle

\section{Introduction} \label{PartI}
\noindent In this paper we focus on the role of thermal pressurization as the main culprit behind frictional stress drop (apparent strain softening) under large coseismic slip. We do so by considering the energy, mass moment, angular moment balance and Thermo-Hydro-Mechanical (THM) couplings, that account for the friction drop during coseismic slip (\cite{Rattez2018b,Rattez2018a}). The THM couplings act on the region inside the fault that accumulates the majority of coseismic slip. This region has a thickness of some milimeters and is called the fault gouge (see \cite{myers2004evolution,Sibson2003}). The problem of shearing of the fault gouge in the presence of THM couplings was first analyzed in \cite{lachenbruch1980frictional} using a classical Cauchy continuum in the case of homogeneous deformation inside the fault gouge. However, the stability of the proposed homogeneous solution is not guaranteed and growing perturbations of the plastic strain field are possible, due to the apparent softening introduced to the model from thermal pressurization. 
This leads to strain localization and mesh dependent results in the case of finite element analyses.\\
\newline
\noindent \cite{Rice2006}, expanding on the solution of \cite{Mase1987}, presented a solution to the above problem for a strain localization profile concentrated on a plane of zero thickness by taking into account the mass and energy balance equations for the THM couplings. Through the use of a strain-rate hardening friction law, it was later shown by \cite{rice2014stability} and \cite{platt2014stability} that the solution for a localization profile of finite thickness lies between the solutions of uniform slip (see  \cite{lachenbruch1980frictional}) and slip on a mathematical plane (see \cite{Mase1987,Rice2006}). In \cite{Rice2014,Platt2014} the solution is shown to approach the limiting case of localization on a mathematical plane as 
seismic slip displacement $\delta$ increases.\\
\newline
\noindent The region inside the fault gouge, where strain localizes is called the Principal Slip Zone (PSZ), and is of the order of some hundred micrometers (see \cite{Muhihaus1988,sibson1977fault}). This is where the majority of the stored elastic and potential energies of the fault dissipate, directly affecting the energy budget of the earthquake phenomenon (see \cite{Andrews2005,Kanamori2004}). 
In order to properly model the effect of the THM couplings in the frictional response and energy dissipation inside the fault gouge and the small thickness of the PSZ, we take into account the microstructure of the fault gouge material (see \cite{vardoulakis2018cosserat,Muhihaus1988}).
 This can be done through the modeling of the fault gouge with a higher order micromorphic continuum \cite[see][among others]{Forest2001, germain1973method}. One such continuum is the Cosserat continuum, that introduces characteristic lengths to the problem thus avoiding strain localization on a mathematical plane and mesh dependence in finite elements analyses (see \cite{Rattez2018a,Rattez2018a,sulem2011stability}). In the work of \cite{Rattez2018b,Rattez2018a} the influence of the Cosserat radius and THM couplings in the strain localization width of the PSZ was investigated with the use of linear stability analyses and nonlinear finite element analyses. The nonlinear finite element analyses have shown that apparent softening is increasing while the localization width is decreasing, as the seismic slip velocity increases. However, the investigated slip distance in these analyses was very limited, and only a rate independent constitutive law was used.\\
\newline
\noindent In this paper we expand on the above mentioned works by considering large slip distances and rate dependence. We then investigate the role of seismic slip velocity in the apparent frictional softening and the formation of a shear band (PSZ) inside a fault gouge that is subjected to large coseismic slip observed in earthquakes. Moreover, we investigate both the rate independent and rate dependent cases. In the works of \cite{Mase1987,Rice2006,Rice2014}, the boundary conditions at the fault gouge, were assumed to lie far away and to not influence the frictional response during coseismic slip. However, according to \cite{myers2004evolution},  the fault gouge is surrounded by the host rock, which has different properties that the gouge. In the works of \cite{aydin2000fractures,passelegue2014influence,tanaka2007thermal,yao2016crucial}, the authors observe that the thermal and hydraulic diffussivities of the fault gouge are smaller than the corresponding parameters of the surrounding material (fault walls), by at least one and two orders of magnitude respectively. This difference between the properties of the fault gouge and the fault walls, motivates us to consider that coseismic slip occurs under isothermal drained boundary conditions applied at the boundaries of the fault gouge. We consider, that the boundaries of the fault gouge lie at finite distance, H, equal to the thickness of the fault gouge. Considering the uncertainty of the diffusion parameters between the fault gouge and the fault walls, this is a simplification, and a more accurate description would require the application of more detailed boundary conditions. Experiments that quantify the effect of the boundary conditions in the frictional response of the fault gouge, during large coseismic slip and thermal pressurization need to be performed.\\
\newline
\noindent In the analyses presented in this paper, the seismic slip (in m) is three (3) orders of magnitude larger that the dimensions of the fault gouge (in mm), and large displacements had to be taken into account. Therefore, an adaptive Lagrangian Eulerian method (ALE) was used in order to apply large displacements. We find that after sufficient slip $\delta$ has occurred, the fault gouge tends to regain part of its strength. This frictional regain occurs due to the isothermal drained boundary conditions of the fault gouge. The percentage of the regained strength is dependent on the fault's slip velocity $\dot{\delta}$ as well as the height of the fault gouge. Moreover, our analyses reveal frictional oscillations at the later stages of coseismic slip. These oscillations at the later parts of the analysis are attributed to the existence of a traveling PSZ inside the fault gouge, which in turn indicates the existence of a limit cycle. The analyses agree qualitatively with the recent experimental results of \cite{Badt2020} and can capture important aspects of the empirical rate and state model, which is popular in fault mechanics. In the second part of this paper, see \cite{Alexstathas2022b}, we propose an explanation for our numerical results providing an expansion to the model of thermal pressurization described in \cite{Rice2006}.\\
\newline
\noindent This paper is structured as follows. In section \ref{ch: 5 sec: Problem_description} we proceed with the formulation of the shear band model subjected to large coseismic slip ($\sim$ 1 m). In section \ref{ch: 5 sec: Numerical results} we elaborate further on the effect of the seismic slip velocity on the shear strength of the fault as well as the localization profiles of the principal slip zone. From the non-linear analyses performed, we monitor the evolution of the solution after the onset of bifurcation from the homogeneous displacement field. We notice therefore, a traveling instability inside the medium, which is connected with the appearance of a limit cycle (\cite{strogatz-2000s}) in the later stages of the analysis. This limit cycle is responsible for the oscillatory behavior of the fields inside the fault gouge. This behavior naturally enhances the frequency content of the near fault earthquake spectra (see \cite{Aki1967a, BRUNEJN1970, Haskell1964, Tsai2020}), since the frictional response is no longer monotonically decreasing. The traveling shear band instability discussed in this paper is also found in metals when similar thermal diffusion mechanisms are considered \cite[see Portevin Le Chatelier phenomenon in][among others]{BENALLAL2006397,Hahner2002,Maziere2010}. In section \ref{ch: 5 viscosity_reference}, we continue our analysis introducing rate dependence in our model through the use of Perzyna type viscoplasticity. This enhances our model with strain rate hardening in the case of velocity stepping, while retaining the strain softening response due to the  THM couplings, at later stages of the analyses. These characteristics of the response are similar to the response of an empirical rate and state friction law \cite[see][]{dieterich1992earthquake,Rice2001,Ruina1983}, largely adopted in the fault mechanics community, without the need of introducing an additional state variable.\\
\newline
\noindent In section \ref{ch: 5 Mase and Smith, Lachenbruch}, we compare our numerical results to the analytical solutions of \cite{lachenbruch1980frictional,Mase1987,Rice2006}. We identify the causes for the difference between the strictly monotonic response predicted by \cite{lachenbruch1980frictional,Rice2006} and the frictional regain observed in our numerical analyses. In Part II \cite{Alexstathas2022b}, we explore further the reasons behind this difference and their implications considering the frictional response and the influence of THM couplings during coseismic slip. Finally, in section \ref{sec: ch: 5 experiments comparison }, we present a comparison of the numerical results of this paper with the recent experimental findings in \cite{Badt2020}, where thermal pressurization was studied in the absence of other weakening mechanisms. These experiments, were done in the same range of parameters as our numerical analyses, and the observed frictional response is in qualitative agreement with our numerical findings.

\section{Problem description \label{ch: 5 sec: Problem_description}}

\subsection{The role of the microstructure in strain localization}
\noindent In this section we summarize the THM equations that govern the behavior of the fault gouge taking into account the role of the microstructure. In a classical Cauchy continuum (also called a Boltzmann continuum, see \cite{eringen1968mechanics}) with a strain softening elastoplastic material, such as in the case of geomaterials, it has been proven in both quasistatic and dynamic regimes that strain localizes in a mathematical plane  (see \cite{vardoulakis1996deformation} among others). This renders the solutions derived from numerical analyses mesh dependent, affecting the amount of the calculated dissipated energy. In order to avoid mesh dependence in our finite element analyses, we will take into account the microstructure. In particular, we will consider that the medium consists of particles with six degrees of freedom, three translations $u_i$ and three rotations $\omega_i,\;i=1,...3$. This corresponds to a first order Cosserat micromorphic continuum (see \cite{vardoulakis2018cosserat}).\\
\newline
\noindent Strain regularization of the corresponding elasto-plastic strain softening medium is of paramount importance. Several researchers have tried to regularize the above problem with the introduction of viscosity effects \cite[see][among others]{wang1997viscoplasticity,needleman1988material}, however, theoretical and numerical analyses in \cite{stathas2022role}, prove that viscous regularization is not capable of regularizing the problem neither in quasi-static or dynamic conditions. Therefore, the only other way of regularizing the problem without postulating an ad-hoc material law is with the use of higher order micromorphic continua such as the Cosserat continuum, which account for the size of the microstructure \cite[see][]{DeBorst1991,forest2001strain,Forest2001,Forest2003,
Muhihaus1988,vardoulakis2018cosserat}.
\subsubsection{Cosserat kinematics}
\noindent We present here, the kinematic description of the Cosserat continuum, starting from the kinematic field of the deformation tensor $\gamma_{ij}$. We define its symmetric part $\gamma_{(ij)}$ as the macroscopic strain $\varepsilon_{ij}$ while its antisymmetric part $\gamma_{[ij]}$ is the difference between macroscopic rotation $\Omega_{ij}$ and the microscopic rotation tensor $\omega_{ij}$. We also take into account the gradient of the microscopic rotation, tensor $\kappa_{ij}$.
\begin{align}
&\gamma_{ij} = \gamma_{(ij)}+\gamma_{[ij]}=u_{i,j}-\omega_{ij}=u_{i,j}+\epsilon_{ijk}\omega_{k},\\
&\gamma_{(ij)}=\varepsilon_{ij} = \frac{1}{2}\left(u^{\prime}_{i,j}+u^{\prime}_{j,i}\right),\\
&\gamma_{[ij]} = \frac{1}{2}\left(u^{\prime}_{i,j}-u^{\prime}_{j,i}\right)=\frac{1}{2}\left(u_{i,j}-u_{j,i}\right)-\omega_{ij}=\Omega_{ij}-\omega_{ij},\\
&\kappa_{ij}=\omega_{i,j},
\label{ch: 5 Cosserat_relations}
\end{align}
where $\epsilon_{ijk}$ is the Levi-Civita permutation tensor.

\subsubsection{Linear and angular momentum balance equations}

\noindent As is the case with Cosserat strains $\gamma_{ij}$, the Cosserat stress tensor $\tau_{ij}$ is also not symmetric. The gradient of micro rotations introduces also Cosserat moments (also called couple stresses) $\mu_{ij}$ to the balance equations. In contrast to Cauchy continua, $\tau_{ij}$ can be decomposed into a symmetric, $\tau_{(ij)}=\sigma_{ij}$, and a non-zero antisymmetric $\tau_{[ij]}$ part. The balance equations can then be written as:
\begin{align}
&\tau_{ij,j}-\rho\frac{\partial^2 u_i}{\partial t^2} = 0,\nonumber \\
&\mu_{ij,j}-\epsilon_{ijk}\tau_{jk}-\rho I \frac{\partial^2 \omega_i}{\partial t^2} = 0,
\label{ch: 5 Cosserat_balance_eqs}
\end{align}
where $\rho\;\text{and}\; I$ are, respectively, the density and microinertia, which are considered isotropic here.

\subsection{Energy balance equation}
\noindent The conservation of energy in a quasi-static transformation, where the material yields producing heat in the form of plastic work, namely dissipation, is expressed, assuming Fourier's law, as:
\begin{align}
&\rho C \left(\frac{\partial T}{\partial t}-c_{th} T_{,ii}\right) = \sigma_{ij}\dot{\epsilon}^p_{ij}+\tau_{[ij]}\dot{\gamma}^p_{[ij]}+\mu_{ij}\dot{\kappa}^p_{ij},
\label{ch: 5 Energy_balnca_eq}
\end{align}
\noindent where $c_{th}=\frac{k_T}{\rho C}, k_T $ are defined as the thermal diffusivity and thermal conductivity of the medium respectively, and $\dot{\epsilon}^p_{ij},\dot{\gamma}^p_{ij},\dot{\kappa}^p_{ij}$ are the rates of the symmetric and antisymmetric part of the plastic strain, and plastic curvature tensors respectively. We neglect the advective derivative, since the porosity of the solid skeleton, $\chi$, of the fault gouge is very small, expecting to lead to small fluid velocities resulting from Darcy's law, \cite[see][for the full derivation]{Rattez2018a}. 

\subsection{Mass balance equation}
\noindent In the case of porous media as the one discussed here, the medium consists of both a fluid phase and a solid phase (insoluble to the fluid), which we consider to communicate perfectly in whole. Meaning no effects of tortuosity and no distinction between principal and secondary pore fluid network will be taken into account. The two phases communicate with each other by acting forces to one another due to different deformation properties \cite[see][]{coussy2004poromechanics, puzrin2001thermodynamics,stefanou2016cosserat}. Finally, the local form of the mixture mass balance equation is given according to \cite{Rattez2018a}:
\begin{align}
\frac{\partial p}{\partial t} =& c_{hy}p_{,ii}+\frac{\lambda^*}{\beta^*}\frac{\partial T}{\partial t}-\frac{1}{\beta^*}\frac{\partial \epsilon_v}{\partial t},
\label{ch: 5 mass_balance_main}
\end{align} 
\noindent where $c_{hy}=\frac{\chi}{\eta^f \beta^*}$ is the hydraulic diffusivity, expressed with the help of the porosity of the solid skeleton $\chi$ and the pore fluid viscosity $\eta^f$, while $\beta^*=n\beta^f+(1-n)\beta^s$, $\lambda^*=n\lambda^f+(1-n)\lambda^s$ are the mixture's compressibility and expansivity respectively \cite[see][]{vardoulakis1986dynamic}. Finally, $\beta^{(s,f)}$ and $\lambda^{(s,f)}$ are the compressibilities and thermal expansivities per unit volume of the respective fluid and solid phase.\\ 
\newline
\noindent During shearing of a fault, friction at the principal slip zone (PSZ) is responsible for the dissipation of the elastic unloading energy into heat. The plastic work produced that way contributes to the energy equation \eqref{ch: 5 Energy_balnca_eq}. Temperature increase leads to pressure increase according to mass balance equation \eqref{ch: 5 mass_balance_main}. In what follows the Terzaghi theory of effective stress is assumed to hold true.
\subsection{Cosserat thermo-elastoplasticity}
\noindent The general constitutive equations in elasticity for a centrosymmetric Cosserat material relating stresses and Cosserat moments to Cosserat strains and curvatures are given by \cite{vardoulakis2018cosserat}:
\begin{align}
\tau_{ij} =& C^{e}_{ijkl}\gamma_{kl}, \nonumber\\ 
\mu_{ij} =& M^{e}_{ijkl}\kappa_{kl}.
\end{align}
The elastic stiffness tensors $C^e_{ijkl}, M^e_{ijkl}$ are derived from 
\begin{align}
C^e_{ijkl} =& \left(K-\frac{2}{3}G\right)\delta_{ij}\delta_{kl}+\left(G+G_c\right)\delta_{ik}\delta_{jl}+\left(G-G_c\right)\delta_{il}\delta_{jk},\\
M^e_{ijkl} =& \left(L-\frac{2}{3}M\right)\delta_{ij}\delta_{kl}+\left(M+M_c\right)\delta_{ik}\delta_{jl}+\left(M-M_c\right)\delta_{il}\delta_{jk}.
\end{align}
\noindent We notice that additionally to the elastic moduli used by the Cauchy media $(K,G)$ denoting isotropic compression and shear moduli respectively, four additional constants are added $G_c, L,M, M_c$ referring to the anti-symmetric part of Cosserat deviatoric stresses, the spherical part of Cosserat moments, the symmetric and anti-symmetric deviatoric parts of the Cosserat moments respectively. The rate independent elastoplastic constitutive relations for the coupled THM problem are given as follows:
\begin{align}
\label{ch: 5 Constitutive_relation_thermo_elasto_plastic_final_main}
&\dot{\tau}_{ij}=C^{ep}_{ijkl}\dot{\gamma}_{kl}+D^{ep}_{ijkl}\dot{\kappa}_{kl}+E^{ep}_{ijkl}\dot{T}\delta_{kl}\\
&\dot{\mu}_{ij}=M^{ep}_{ijkl}\dot{\kappa}_{kl}+L^{ep}_{ijkl}\dot{\gamma}_{kl}+N^{ep}_{ijkl}\dot{T}\delta_{kl}.
\end{align}
\noindent The superscript $(^{ep})$ denotes the elastoplastic matrices during loading, whose detailed expressions are given in Appendix \ref{Appendix D}. 
\subsection{Large displacements}
\noindent Since our analyses reach displacements far greater than the 1D model's geometrical dimensions, we need to take into account large changes in the volume of the element along with rotations of the reference frame. Our application involves pure shearing of the fault gouge layer and therefore, the displacement derivatives with respect to $x_1$ axes are zero (see Figure \ref{ch: 5 fig:1D Cosseral_model}). We notice that the displacement parallel to the $x_2$ direction is expected to be small. This is because no additional loading will be applied in the vertical direction during shearing, while from the plastic potential we have that for a mature fault the dilatancy angle is very low $\beta \sim 0$ as discussed in \cite{Rice2006,Sulem2016}. The thermal expansion and compressibility coefficients are also very small so that in the observed temperature and pressure range their effects are minimal. Therefore, the deformation tensor $F_{ij}$ can be written in matrix form as: 
\begin{align}
F=
\begin{bmatrix}
\frac{\partial x_1}{\partial X_1} & \frac{\partial x_1}{\partial X_2}\\
\frac{\partial x_2}{\partial X_1} & \frac{\partial x_2}{\partial X_2}\\
\end{bmatrix}
=
\begin{bmatrix}
1+\frac{\partial u_1}{\partial X_1} & \frac{\partial u_1}{\partial X_2}\\
\frac{\partial u_2}{\partial X_1} & 1+\frac{\partial u_2}{\partial X_2}\\
\end{bmatrix}
\approx
\begin{bmatrix}
1 & \frac{\partial u_1}{\partial X_2}\\
0 & 1\\
\end{bmatrix},
\end{align}
where $X_i$, $x_i$ are the reference and current configuration coordinates, respectively.
From the above we establish that $\det{F}\approx1$. Therefore no large volume changes are expected to take place during the shearing phase of the analysis. This conclusion is supported also by the numerical findings in which the volumetric strain is adequately small $\epsilon_v<0.005$. To account for any effects that large displacements may introduce to our model we have also run a series of analyses based on an Arbitrary Lagrangian Eulerian (ALE) method \cite[see][]{Donea2004}, where at every iteration we update the new mesh position. The change in the mesh is kept at every converged increment otherwise the cumulative change inside the increment is discarded and the procedure starts anew.\\
\newline
\noindent The question of the plastic work due to large rotations of the microstructure can be covered with the help of the ALE approach considering an additive decomposition of the curvature tensor into an elastic and a plastic part. A more general description of the Cosserat continuum in elasto-plasticity under large deformations can be found in \cite{Forest2003,forest2020continuum}. There, the authors adopt the multiplicative decomposition for the deformation tensor $F_{ij}$ into the elastic and plastic parts $F^e_{ij},\;F^p_{ij}$, while again the additive decomposition for the curvature tensor $\kappa_{ij}$ is pursued.  

\subsection{Normalized system of equations. \label{ch: 5 Normalized_system_of_equations}}
\noindent Equations \eqref{ch: 5 Cosserat_balance_eqs}, \eqref{ch: 5 Energy_balnca_eq}, \eqref{ch: 5 mass_balance_main} constitute the nonlinear system that describes the behavior of the fault. We define the following dimensionless parameters, $\bar{x}=\frac{x}{\text{H}_0},\;\bar{t}=\frac{t}{t_0}\;\bar{u}_i=\frac{u_i}{u_0},\;\bar{\tau}_{ij}=\frac{\tau_{ij}}{\tau_0},\;\bar{\mu}_{ij}=\frac{\mu_{ij}}{\mu_0},\;\bar{T}=\frac{T}{T_0},\;\bar{p}=\frac{{p}}{\tau_0}$, where $\text{H}_0,\;t_0,\;\tau_0,\;\mu_0,\;T_0,\;p_0$ are characteristic length, time, stress, moment, temperature and pressure quantities, respectively. Furthermore, we note that there are specific relations between the characteristic moment $\mu_0,\;\text{H}_0,\;\tau_0$ based on their dimensions i.e. $\mu_0=\tau_0 \text{H}_0$.\\
\newline
\noindent The non-linear, normalized equations of the problem are given then as:
\begin{align}
&\bar{\tau}_{ij,j}-I_1\frac{\partial^2 \bar{u}_i}{\partial \bar{t}^2}=0,\nonumber\\
&\bar{\mu}_{ij,j}-\epsilon_{ijk}\bar{\tau}_{jk}-I_2 \frac{\partial^2\omega_i}{\partial \bar{t}^2}=0,\nonumber\\
&\frac{\partial\bar{T}}{\partial \bar{t}}=\frac{c_{th}t_0}{\text{H}^2_0}\frac{\partial^2 \bar{T}}{\partial \bar{x}^2}-\frac{\tau_0}{T_0}\left(\bar{\sigma}_{ij}\dot{\varepsilon}_{ij}+\bar{\tau}_{ij}\dot{\gamma}_{ij}+\bar{\mu}_{ij}\dot{\kappa}_{ij}\right),\nonumber\\
&\frac{\partial \bar{p}}{\partial t}=\frac{c_{hy} t_0}{\text{H}^2_0}\frac{\partial^2 \bar{p}}{\partial \bar{x}^2}+\frac{\lambda^*}{\beta^*}\frac{T_0}{p_0}\frac{\partial \bar{T}}{\partial \bar{t}}-\frac{1}{\beta^*p_0}\frac{\partial \varepsilon_v}{\partial \bar{t}}, \label{ch: 5 normalized_system_1}
\end{align}
\noindent where, $I_1=\rho\frac{u_0 \text{H}_0}{t^2_0\tau_0}$ and $I_2=\frac{\rho I}{t^2_0 \tau_0}$.
\noindent We consider the following characteristic dimensions and their relations, in order to investigate properly the effect of each term in the behavior of the system:
\begin{align}
\text{H}_0=u_0=\text{H},\;\tau_0=p_0=\sigma_n-p^{init},\;T_0=\frac{(\sigma_n-p^{init})\beta^*}{\lambda^*},\;t_0=\frac{\text{H}}{V},
\end{align} 

\noindent where H is the height of the fault gouge layer, $\sigma_n,\;p^{init}$ are the normal stress (constant during the analysis) and the initial pore fluid pressure, respectively, and $V$ is the constant shear velocity applied at the boundaries of the layer. Here we emphasize that different scaling parameters may be chosen for the non dimensionalization of the system. In particular in \cite{sulem2011stability} the authors chose to scale time with the help of the thermal diffusivity ($t_0=\frac{R^2}{c_{th}}$). Another candidate for time non dimensionalization are the characteristic timescales found in the homogeneous shear solution defined by \cite{lachenbruch1980frictional} or the shear on a plane defined by \cite{Mase1987}. These solutions involve parameters such as the diffusivities $c_{th},\;c_{hy}$ in the normalization of time. However, as explained in \cite{Rice2006,Rice2014,Platt2014}, these quantities can significantly change during shearing of the fault gouge.\\
\newline
\noindent The proposed scaling used here ($\tau_0=\frac{\text{H}}{V}$), does not contain the diffusion parameters itself, and therefore, it keeps the inertia effects independent of the diffusion parameters of the system. Application of this scaling in the system of equations
 \eqref{ch: 5 normalized_system_1}, indicates the influence of the layer's height, and the shearing rate in the numerical analyses. We will make use of this later in section \ref{ch: 5 Traveling_instability_new}. We observe that the height of the layer influences the diffusion terms of the system \eqref{ch: 5 normalized_system_1}, and the rotational inertia of the microstructure. Increase of the layer's height $\text{H}$, decreases the efficiency of the diffusion terms, further intensifying thermal pressurization.\\
 \newline 
\noindent Based on the proposed scaling the inertia terms are given by: $I_1=\frac{\rho V^2}{\tau_0}=0.15\;10^{-3}\ll 1,\;I_2=\frac{ I V^2}{\rho\tau_0\text{H}^2}=\frac{2}{5}\frac{\rho}{\tau_0} \left(\frac{R V}{\text{H}}\right)^2=6\;10^{-9}\ll 1$. This indicates that the inertia terms can be neglected during coseismic slip. \cite{platt2014stability,rice2014stability} investigated the role of inertia in the localization width of the principal slip zone in the constant seismic slip velocity analyses. They concluded that inertia does not significantly affect the width of the localized zone except at the propagation tip, where the inertial number is significantly high and the localization profiles widen. Additionally, the role of the microstructure (inertia of the grains) was investigated in the linear perturbation analyses in \cite{sulem2011stability}. It is found that for the in situ observed seismic slip velocities up to $\sim 1$ m/s, the localization width does not change significantly compared to the case where inertia is neglected.\\
\subsection{Linear stability analysis}
\begin{figure}[h!]
  \centering
  \includegraphics[width=0.75\textwidth]{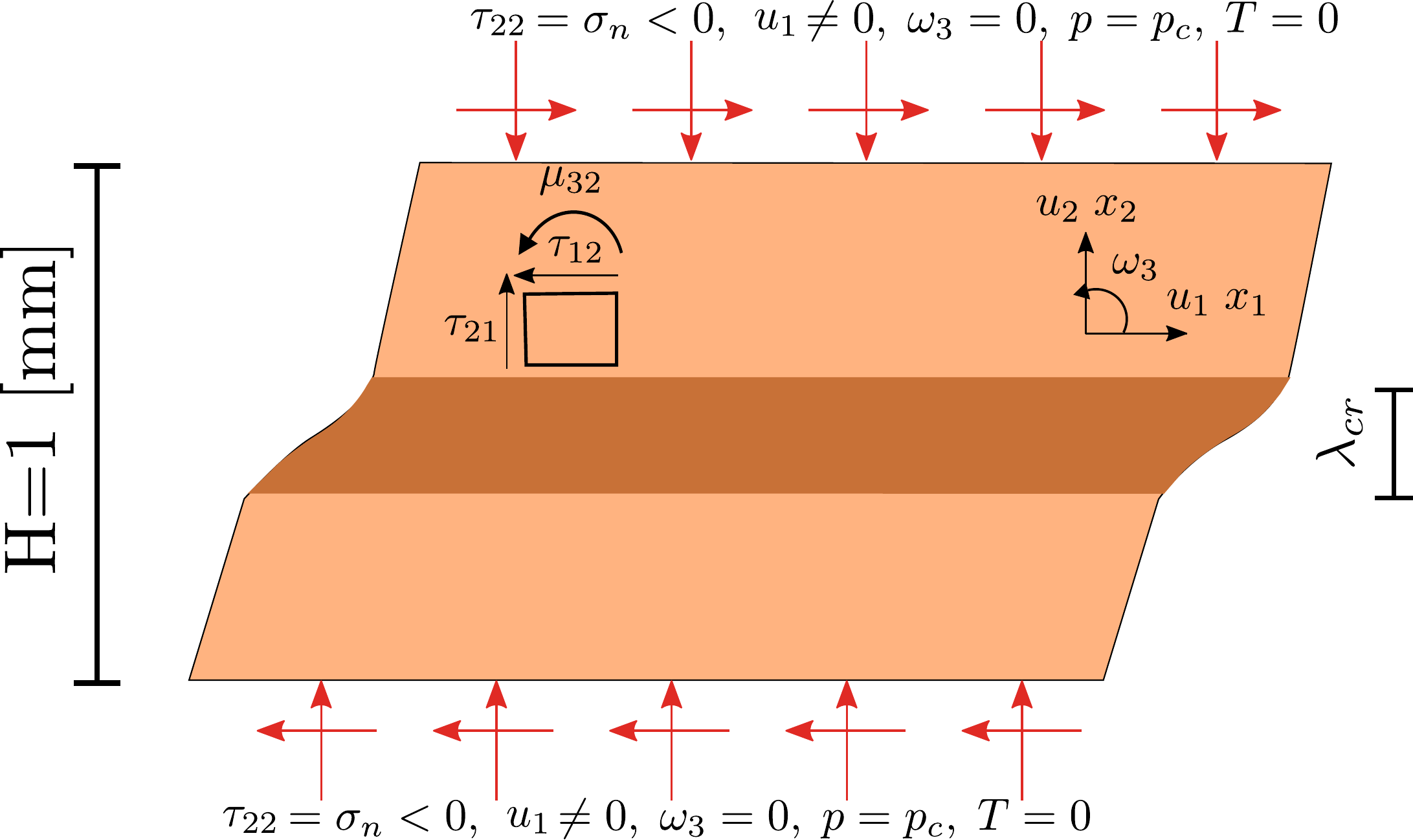}
  \caption{1D consolidated Cosserat layer under shear. }
  \label{ch: 5 fig:1D Cosseral_model}
\end{figure}

\noindent In what follows we refer to the shearing of a 1D layer under constant shear slip velocity at the boundaries as discussed in \cite{lachenbruch1980frictional,rice2006heating}. From now on, in order to reduce notation complexity we remove the $\left(\;\bar{}\;\right)$ sign from the normalized unknowns.
We apply a perturbation $\tilde{\phi}(x_l,t)=[\tilde{\tau}_{ij}(x_l,t),\;\tilde{\mu}_{ij}(x_l,t),\;\tilde{T}(x_l,t),\;\tilde{p}(x_l,t)]$ to the homogeneous solution $\phi^*(t)=[{\tau}^*_{ij}(t),\;{\mu}^*_{ij}(x_l,t),\;{T}^*(t),\;{p}^*(t)]$. Applying the perturbed solution $\phi^*+\tilde{\phi}$ to the above system of equations we obtain the linearized perturbed system,
\begin{align}
&\tilde{\tau}_{ij,j}=0,\nonumber\\
&\tilde{\mu}_{ij,j}-\epsilon_{ijk}\tilde{\tau}_{jk}=0,\\
&\frac{\partial\tilde{T}}{\partial {t}}=\frac{c_{th}t_0}{\text{H}^2_0}\tilde{T}_{,ll}-\frac{\tau_0}{T_0}\left({\sigma}^*_{ij}\tilde{\epsilon}_{ij}+{\tau}^*_{[ij]}\tilde{\gamma}_{[ij]}+\mu_{ij}^*\tilde{\kappa}_{ij}\right),\label{ch: 5 modified energy}\\
&\frac{\partial \tilde{p}}{\partial t}=\frac{c_{hy} t_0}{\text{H}^2_0} \tilde{p}_{,ll}+\frac{\lambda^*}{\beta^*}\frac{T_0}{p_0}\frac{\partial \tilde{T}}{\partial {t}}-\frac{1}{\beta^*p_0}\frac{\partial \tilde{\varepsilon}_v}{\partial {t}}. \label{ch: 5 linearized_system}
\end{align}
\noindent In deriving the equation \eqref{ch: 5 modified energy}, we neglected the perturbation of the stress $\tilde{\tau}_{ij}$ and couple stress tensors $\tilde{\mu}_{ij}$ and considered only perturbations of the plastic deformation and curvature tensors $\tilde{\gamma}_{ij},\tilde{\kappa}_{ij}$. Moreover, we considered that the perturbation in the plastic part of the strain and curvature tensors ($\tilde{\gamma}^p_{ij},\tilde{\kappa}^p_{ij}$) is significantly bigger that the corresponding pertubation of the elastic parts ($\tilde{\gamma}^e_{ij},\tilde{\kappa}^e_{ij}$). Thus due to the additive decomposition of the elastic and plastic parts of the strain and curvature tensors we can assume that ($\tilde{\gamma}_{ij}\approx\tilde{\gamma}^p_{ij},\tilde{\kappa}_{ij}\approx\tilde{\kappa}^p_{ij}$). We inject the constitutive relations of equation \eqref{ch: 5 Constitutive_relation_thermo_elasto_plastic_final_main} into the linearized system \eqref{ch: 5 linearized_system}, expressing the total linearized strains $\tilde{\gamma}_{ij}$ and curvatures $\tilde{\kappa}_{ij}$ with respect to the perturbed displacements $\tilde{u}_i$ and Cosserat rotations $\tilde{\omega}^c_i$ as presented in \cite{Rattez2018b}:
\begin{align}
&C^{ep}_{klmn}(\tilde{u}_{m,nl}+\epsilon_{mnq}\tilde{\omega}^c_{q,l})+E^{ep}_{klmn}\tilde{T}_{,l}\delta_{mn}+D^{ep}_{klmn}\tilde{\omega}^c_{m,nl}-\tilde{p}_{,l}\delta_{kl}=0,\\
&M^{ep}_{klmn}\tilde{\omega}^c_{m,nq}+L^{ep}_{klmn}(\tilde{u}_{m,nq}+\epsilon_{mnq}\tilde{\omega}^c_{q,l})-\epsilon_{klm}(C^{ep}_{lmnq}(\tilde{u}_{n,q}+\epsilon_{nqr}\tilde{\omega}^c_r)\nonumber\\
&+E^{ep}_{lmnq}\tilde{T}\delta_{nq}+D^{ep}_{lmnq}\tilde{\omega}^c_{n,q}-\tilde{p}\delta_{lm})=0,\\
&\frac{\partial\tilde{T}}{\partial {t}}=\frac{c_{th}t_0}{\text{H}^2_0}\tilde{T}_{,ll}-\frac{\tau_0}{T_0}\left({\tau}^*_{ij}(\tilde{u}_{i,j}-\epsilon_{ijk}\tilde{\omega}_k)+\mu_{ij}^*\tilde{\omega}_{i,j}\right),\label{ch: 5 perturbed_temp}\\
&\frac{\partial \tilde{p}}{\partial t}=\frac{c_{hy} t_0}{\text{H}^2_0}\tilde{p}_{,ll}+\frac{\lambda^*}{\beta^*}\frac{T_0}{p_0}\frac{\partial \tilde{T}}{\partial {t}}-\frac{1}{\beta^*p_0}\frac{\partial \tilde{u}_{k,k}}{\partial {t}}.
 \label{ch: 5 full_linearized_system}
\end{align}
\noindent
We assume shearing under constant shear slip velocity $\dot{\delta}$, and therefore, the layer is sheared under linear in time Dirichlet boundary conditions. Moreover, we assume shearing of the layer under isothermal and drained boundary conditions for the temperature and pressure diffusion equations. We introduce a perturbation of the form $[\tilde{u}_i,\tilde{\omega}_i,\tilde{T},\tilde{p}]=[u_0,\omega_0,T_0,p_0]\exp(s\bar{t})\exp(ik\bar{x}_jn_j),\;k=\frac{2\pi}{\lambda}$, where $\lambda=\frac{h}{2\pi N}$, where $N$ is an integer satisfying the boundary conditions.\\
\newline
\noindent In this paper we are mainly interested in the shearing of a 1D layer (see Figure \ref{ch: 5 fig:1D Cosseral_model}). In this context only the derivatives along the $x_2$ axis survive, therefore, the above system is reduced as follows:
\begin{align}
&C^{ep}_{2222}\tilde{u}_{2,22}+D^{ep}_{2232}\tilde{\omega}^c_{3,22}+E^{ep}_{2222}\dot{T}_{,2}-\tilde{p}_{,2}=0,\\
&C^{ep}_{1212}(\tilde{u}_{1,22}+\epsilon_{123}\tilde{\omega}^c_{3,2})+D^{ep}_{1232}\tilde{\omega}^c_{3,22}=0,\\
&M^{ep}_{3232}\tilde{\omega}^c_{3,22}+L^{ep}_{3212}(\tilde{u}_{1,22}+\epsilon_{123}\tilde{\omega}^c_{3,2})+C^{ep}_{2112}(\tilde{u}_{1,2}+\tilde{\omega}^c_3)-C^{ep}_{1221}(\tilde{u}_{2,1}-\tilde{\omega}^c_3)\nonumber\\
&+D^{ep}_{2132}\tilde{\omega}^c_{3,2}+D^{ep}_{1232}\tilde{\omega}^c_{3,2}=0,\\
&\frac{\partial\tilde{T}}{\partial {t}}=\frac{c_{th}t_0}{\text{H}^2_0}\tilde{T}_{,22}-\frac{\tau_0}{T_0}\left({\tau}^*_{21}(-\tilde{\omega}_3)+{\tau}^*_{12}(\tilde{u}_{2,1}+\tilde{\omega}_3)+\mu_{32}^*\tilde{\omega}_{3,2}\right),\\
&\frac{\partial \tilde{p}}{\partial t}=\frac{c_{hy} t_0}{\text{H}^2_0}\tilde{p}_{,22}+\frac{\lambda^*}{\beta^*}\frac{T_0}{p_0}\frac{\partial \tilde{T}}{\partial {t}}-\frac{1}{\beta^*p_0}\frac{\partial \tilde{u}_{2,2}}{\partial {t}}.
 \label{ch: 5 full_linearized_system}
\end{align}
\qquad
\subsection{Traveling instabilities \label{ch: 5 Traveling_Instabilities}}
\begin{figure}[h!]
  \centering
  \includegraphics[width=0.9\textwidth]{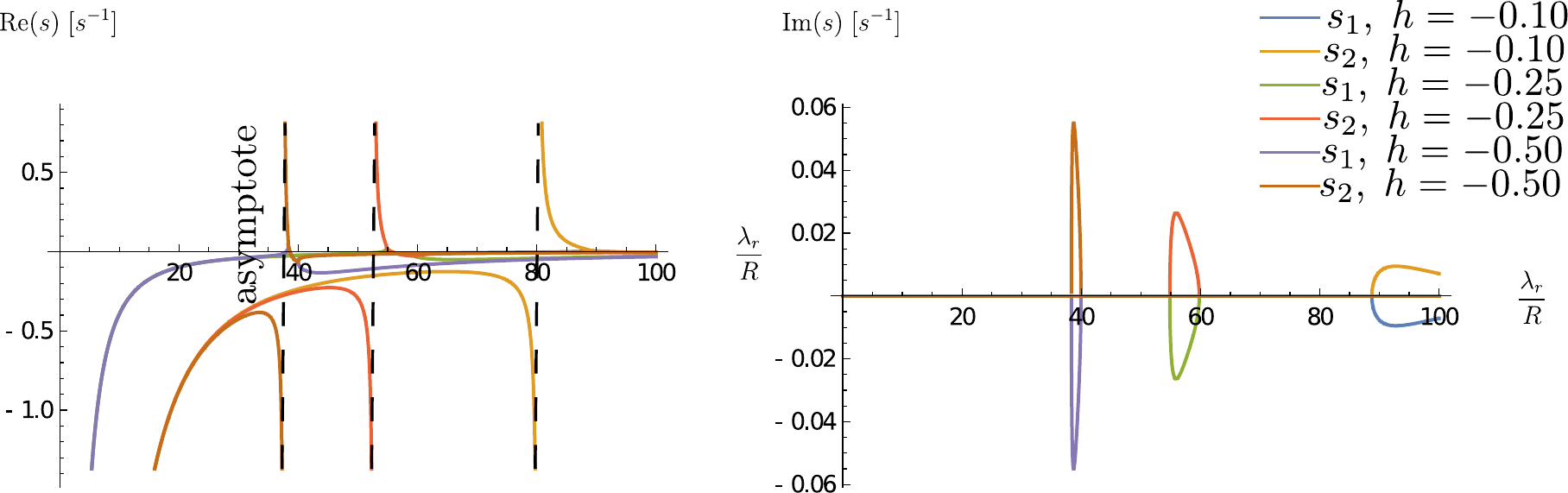}
  \caption{On the left: Real part of the two roots of the characteristic polynomial for different values of the hardening parameter h. There is always a critical wavelength value $\lambda_{cr}$ greater than zero, for which the Lyapunov exponent tends to infinity. On the right: Imaginary parts of the roots for the same softening values. As softening intensifies, positive real values of $s$ with non zero imaginary part appear on a $\lambda_r\neq 0$, leading to a traveling perturbation of a non zero critical wavelength.}
  \label{ch: 5 fig: Lyapunov coefficient}
\end{figure}

\begin{figure}[h!]
  \centering
  \includegraphics[width=0.9\textwidth]{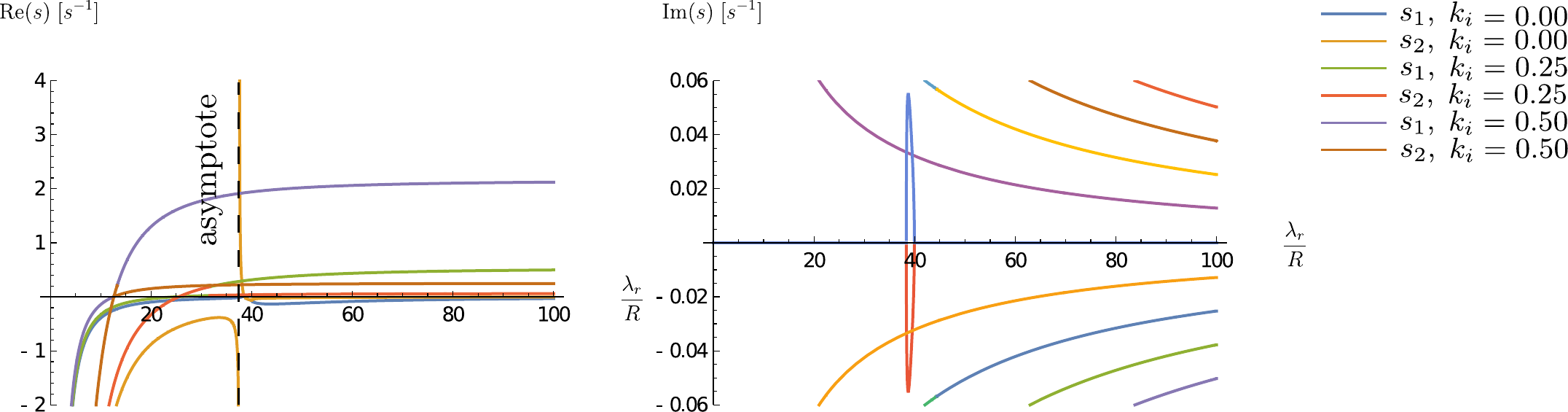}
  \caption{On the left: Real part of the two roots of the characteristic polynomial for different values of the attenuation coefficient $k_i$. The two roots start negative and eventually they pass to positive values for the real part of $s$. The wavelength each root changes its sign as well as the maximum value reached are characteristics of the value of $k_i$. On the right: Imaginary parts of the roots for the same softening values and values of the attenuation coefficient $k_i$. For $k_i \neq 0$, the magnitude of the imaginary part of the roots Im[$s_1,s_2$] starts from $\infty$ when $\lambda_r=0$ and slowly attenuates. This is in high contrast to the behavior exhibited when $k_i=0$. The perturbation increasing the fastest is the one lying on $k_i=0$.}
  \label{ch: 5 fig: Lyapunov coefficient different ki}
\end{figure}

\noindent In this section, we advance beyond the linear stability analyses carried out in \cite{Rattez2018a}, as we are mainly concerned with the behavior of the imaginary part of the complex eigenvalues defining the Lyapunov coefficient $s$. In the context of Lyapunov stability analysis the eigenvalues with positive real part show that the system is unstable. Moreover, if the eigenvalue in question is complex then the instability is characterized as a flutter instability (see \cite{Brauer1969}). In partial differential equations though, the appearance of imaginary parts in $s$ is associated to the nucleation of traveling waves as cited in \cite{rice2014stability,platt2014stability} and presented in \cite{stathas2022role}. More specifically, by assuming a perturbation $\tilde{\phi}=[\tilde{u}_i,\tilde{\omega}_i,\tilde{T},\tilde{p}]$ of complex Lyapunov coefficient $s=s_r+s_i i$ and complex wavenumber $k=k_r+k_i i$ of the form:
\begin{align}
&\tilde{\phi} =\phi_0\exp(s_r\bar{t}+k_ix)\exp[i(s_it+k_r\bar{x})],\label{ch: 5 perturbation_developped}
\end{align}
\noindent where, $\phi_0=[u_0,\omega_0,T_0,p_0]$, we observe that traveling perturbations appear in the medium, as the second factor or equation \eqref{ch: 5 perturbation_developped} becomes a sinusoidal.\\
\newline
\noindent The existence of the traveling perturbations together with the special kind of drained, isothermal, Dirichlet (essential) boundary conditions of the PDE system leads to reflection of the traveling perturbation near the fault gouge boundaries and may lead to the appearance of a limit cycle. 
We note here that the applied boundary conditions are extremely important for the behavior of the instability. Contrary to the previous case of Dirichlet conditions, In the case of adiabatic undrained (Neumann) boundary conditions, the spatial profiles of pressure and temperature do not allow for a reflection of the traveling instability at the boundaries and therefore, the perturbation stations in one of the boundaries of the model. The existence of a limit cycle and the traveling instability is a characteristic similar to the Portevin Le Chatelier effect found in metals \cite[see][among others]{Hahner2002,Maziere2010,wang1997viscoplasticity}. 
\noindent In the case of metals, the identified limit cycle, is attributed to the diffusion mechanisms present in the medium and the inherent coupling between balance and diffusion equations (see \cite{BENALLAL2006397,BENALLAL20081916,Hahner2002,Maziere2010}).\\
\newline
\noindent 
\noindent We note that from the linear stability analysis of the previous system around the solution of homogeneous deformation (see \cite{sulem2011stability}), the dominant perturbation does not have a complex Lyapunov coefficient. This, however, is not necessarily true around a localized solution away form the initial homogeneous deformation. In the last case the influence of the apparent softening due to thermal pressurization needs to be taken into account. Assuming a correspondence exists between the apparent softening and a mechanical softening parameter, we expect flutter instabilities to be present once the system starts exhibiting a softening behavior.\\ 
\newline
\noindent In Figure \ref{ch: 5 fig: Lyapunov coefficient}, we present the effect of the softening parameter $h$ on the localization length $\lambda_r$ of the maxima of the real and imaginary parts $(s_r,s_i)$ of the Lyapunov coefficient $s$. We note that the value of the real part of Lyapunov coefficient $s_r$ is negative for perturbations of zero wavelength, therefore no localization on a mathematical plane can occur, independently of the softening parameter $h$. This is to be expected in the case of a Cosserat continuum due to the introduction of an internal length, which regularizes the problem. Initially for small values of softening the wavelengths corresponding to the maxima of $s_r,s_i$ do not match. The wavelength corresponding to the growing perturbation exhibits an unbounded positive real part and zero imaginary part for the Lyapunov coefficient. This means that at the initial stages of coseismic slip, the strain localization is not traveling in the medium. However, as yielding continues and thermal pressurization leads to more pronounced softening, the imaginary part catches up with the real part, meaning that both the maxima of the real and imaginary parts of the Lyapunov coefficient $s$ occur for the same perturbation wavelength. This leads to a traveling strain localization inside the layer (see Figure \ref{ch: 5 fig: Lyapunov coefficient} and section \ref{ch: 5 sec: Numerical results}). 
\\
\newline
\noindent For the value of the softening parameter $h_s=-0.50$ that produces the traveling perturbation of finite width, we further examine the effect of a non negative imaginary part in the wavenumber $k=k_r+k_i i$. The imaginary part of $k$ is responsible for the change of the perturbation amplitude with space. In a bounded problem such as the one discussed here, its effect is not important as the amplitude will remain bounded with distance. It is of interest, however, to examine whether traveling modes of strain localization are possible in the more general context. For a given value of the softening parameter $h_s$ and treating $k_i$ as a parameter, the plots of the real and imaginary part of the roots of the characteristic polynomial $s_{1,2}$ are shown in Figure \ref{ch: 5 fig: Lyapunov coefficient different ki}. In the left part of Figure \ref{ch: 5 fig: Lyapunov coefficient different ki}, we notice that localization on a mathematical plane is avoided for any $k_i$, since for $\lambda_r=0$ the real part of the Lyapunov coefficient $s$ tends to $-\infty$. This is in contrast with viscous regularization, where it was recently shown that strain localization on a mathematical plane is always possible (see \cite{stathas2022role}). This behavior is owed to the internal length introduced in the Cosserat continuum that regularizes effectively strain localization.\\
\newline
\noindent On the right part of Figure \ref{ch: 5 fig: Lyapunov coefficient different ki} we notice that the behavior of the imaginary part of the roots $s_1,s_2$ is symmetric around $k_i=0$, as expected. Therefore, we focus our attention to values of $k_i>0$. The roots present a positive real part for a range of values of $\lambda_r$ (see Figure \ref{ch: 5 fig: Lyapunov coefficient different ki}). For $k_i \neq 0$ the positive real part of the Lyapunov coefficient, $s_r=\text{Re}[s]$, is bounded and obtains its maximum value as $\lambda_r$ tends to $\infty$. 
For $k_i \neq 0$, the maximum value of the imaginary part of $s_{1,2}$ is obtained for $\lambda_r \rightarrow 0$ and then it slowly attenuates as $\lambda_r \rightarrow \infty$. Thus when $k_i \neq 0$ the uniform strain profile dominates. Finally, when $k_i=0$, we note again that Re$[s]\to +\infty$ and Im$[s]\neq 0 $ at a finite value of $\lambda_r$. We conclude, from the above, that traveling perturbations of unbounded increasing amplitude and finite width are possible in the case $(s_r>0, s_i\neq 0, k_i = 0)$. 
\section{Numerical Analyses \label{ch: 5 sec: Numerical results}}
\noindent A 1D model of a Cosserat layer was used, where shear displacement was applied to the boundaries of the layer, while rotations were blocked at both ends. Figure \ref{ch: 5 fig:1D Cosseral_model} describes the model in more detail. The layer was discretized using 80 finite elements, with quadratic shape functions for the displacement field $u_i$ and linear shape functions for the rotations $\omega_i$. Reduced integration scheme was used for the displacement field compared to full for the rotation field \cite[see][]{godio2016multisurface}. These element parameters were taken as a result of a mesh convergence investigation of different shape functions and number of Gauss points that was performed in a previous work in order to find the optimal mesh description \cite[see][]{stathas2019stefanou}.
The mesh characteristics are summarized in Table \ref{ch: 5 table:mesh_properties}. The Cosserat material properties used to describe a mature fault in the seismogenic zone are summarized in Table \ref{ch: 5 table:material_properties}, where a relatively high value for the friction coefficient $\mu$ has been used with respect to the values provided in \cite{Rice2006,Rempel2006,Rice2014} and path averaged values for $\lambda^*$, $\beta^*$ were considered, as proposed in \cite{rice2006heating,Rice2014}.
\begin{table}[h!] 
\begin{center}
\begin{tabular}[]{l l l}
\hline
 & $u_i$ & $\omega_i$  \\
\hline
\hline
Element type & Quadratic & Linear \\
Integration scheme & Reduced & Full\\
Number of elements &\multicolumn{2}{c}{80}\\
\hline
\end{tabular}
\caption{Mesh properties of the problem.}
\label{ch: 5 table:mesh_properties}
\end{center}
\end{table}
\\
\noindent To illustrate the role of seismic slip velocity in the post peak behavior of the fault, we apply two different shear velocity-stepping programs to the model at hand. First we implement a three step procedure described in section \ref{ch: 5 sec: Three_step_procedure} which includes, consolidation of the layer to the stresses and pressure at a depth representative of the seismogenic zone ($7$ km) followed by slow shear of the layer and then by fast shear for a shear slip of 10 mm at each stage. The second program in section \ref{ch: 5 sec: Two_step_procedure} involves initial consolidation and then shear with constant slip velocity, $\dot{\delta}$, ranging from as slow as $0.01$ m/s to $1.0$ m/s for a total of 100 mm of seismic slip $\delta$.
\begin{table}[h!]
\begin{center}
\begin{tabular}[]{l l l l l l}
\hline
Parameters & Values & Properties & Parameters & Values & Properties \\
\hline
\hline
$K$ &$20.\;10^3$ &MPa &$\mu$ & $0.5$ & - \\
$G$ &$10.\;10^3$ &MPa &$\beta$ & $0$ & - \\
$G_c$ &$5.\;10^3$ &MPa &$\lambda^*$ & $13.45\;10^{-5}$ & $/^o$C \\
$L$ &$10^3$ &MPa mm$^2$ &$\beta^*$ & $8.2\;10^{-5}$ & MPa$^-1$ \\
$M$ &$1.5$ &MPa mm$^2$ &$\rho C$ & $2.8$ &MPa/$^o$C \\
$M_c$ &$1.5$ &MPa mm$^2$ &$c_{hy}$ & $12.0$ &mm$^2$/s$^2$ \\
$R$ &$0.01$ &mm &$c_{th}$ & $1.0$ &mm$^2$/s$^2$ \\
$\sigma_n$ & $200$ &MPa &$\alpha_s$ & $10^{-5}$ &$/^o$C \\
$p_0$ & $66.67$ &MPa &$\chi$ & $12.\;10^{-15}$ &m$^2$ \\
\hline
\end{tabular}
\end{center}
\caption{Material parameters of a mature fault at the seismogenic depth \protect\cite[see][]{rice2006heating,Rattez2018b}.}
\label{ch: 5 table:material_properties}
\end{table}\\
\newline
\noindent A second series of analyses were also run, where the seismic slip displacement is set to 1 m and the seismic slip velocity to $1$ m/s - typical values observed in nature during large coseismic slip. All these analyses go far beyond the previous limit of $5$ mm presented in \cite{Rattez2018b,rattez2018numerical}, and allow us to observe new and interesting phenomena. The higher seismic slip displacement, allows us a deeper understanding of the phenomenon of thermal pressurization since it is shown that a traveling instability is formed inside the gouge due to the existence of a limit cycle in later parts of the analysis (see Figure \ref{ch: 5 fig: tau_u_velocity_compare_1_ Solutions comparizon}).  This behavior is new compared to previous analyses on the same mechanism of thermal pressurization done with simpler models (see \cite{lachenbruch1980frictional,rice2006heating,rice2014stability}).\\
\newline
\noindent Finally, in section \ref{ch: 5 Traveling_instability_new}, we illustrate here the effect of the boundaries in the traveling velocity $v$ of the PSZ inside fault gouge, by considering two different fault gouge layers of height 1 mm and 2 mm respectively subjected to the same seismic slip displacement $\delta$=1 m with seismic slip velocity $\dot{\delta}$=0.5 m/s. We note, based on the scaled system of equations \eqref{ch: 5 normalized_system_1}, that the two configurations described here, differ only in the diffusion terms. Namely, the thicker layer diffuses pressure and temperature slower, exhibiting more pronounced apparent softening due to thermal pressurization.
\subsection{Shearing of a mature fault gouge under small slip ($\delta=10$ mm) and velocity stepping}\label{ch: 5 sec: Three_step_procedure}
\begin{figure}[h!]
  \centering
  \begin{minipage}{.43\textwidth}
  \centering
  \includegraphics[width=0.9\linewidth]{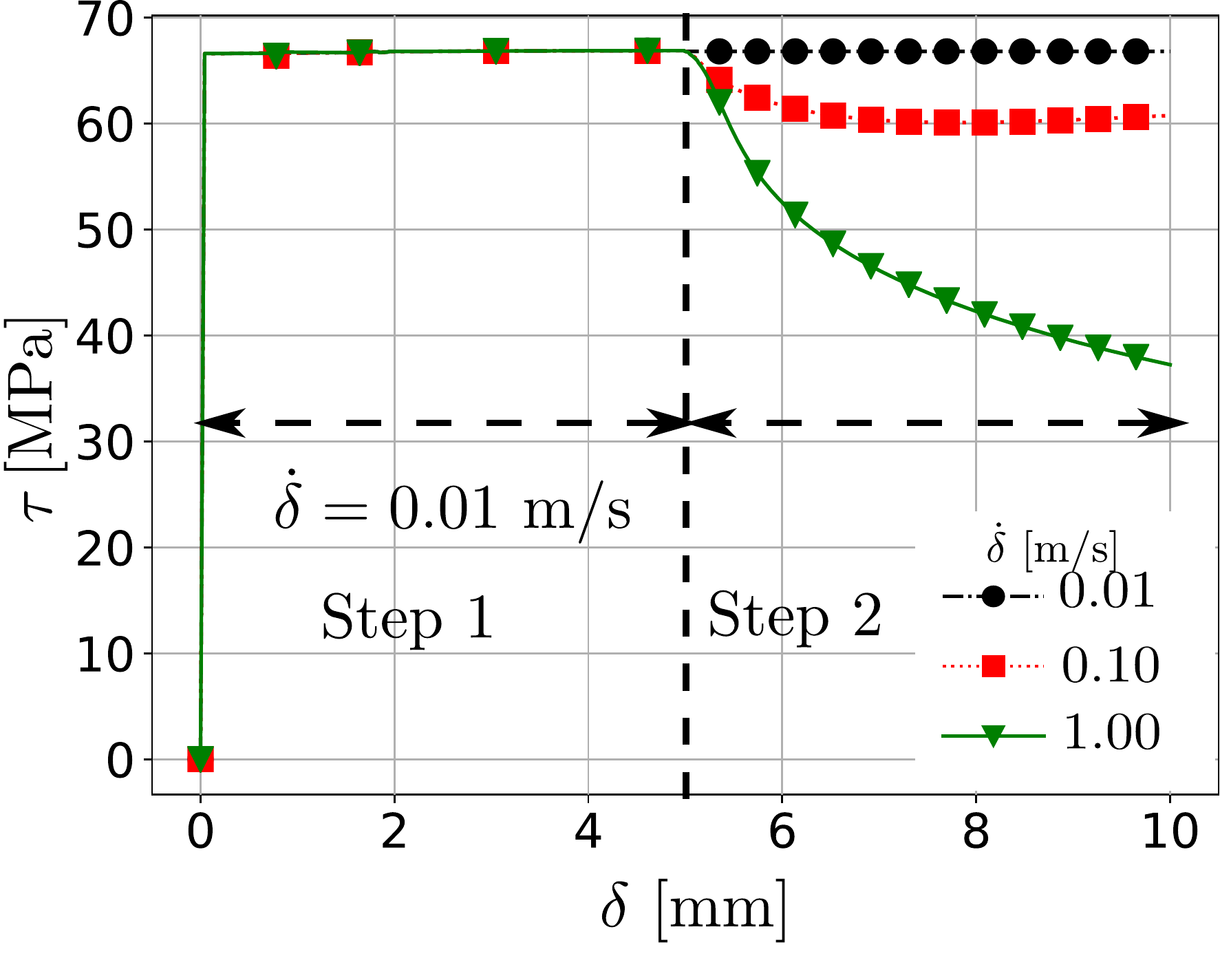}
  \end{minipage}\qquad
  \begin{minipage}{.47\textwidth}
  \centering
  \includegraphics[width=0.9\linewidth]{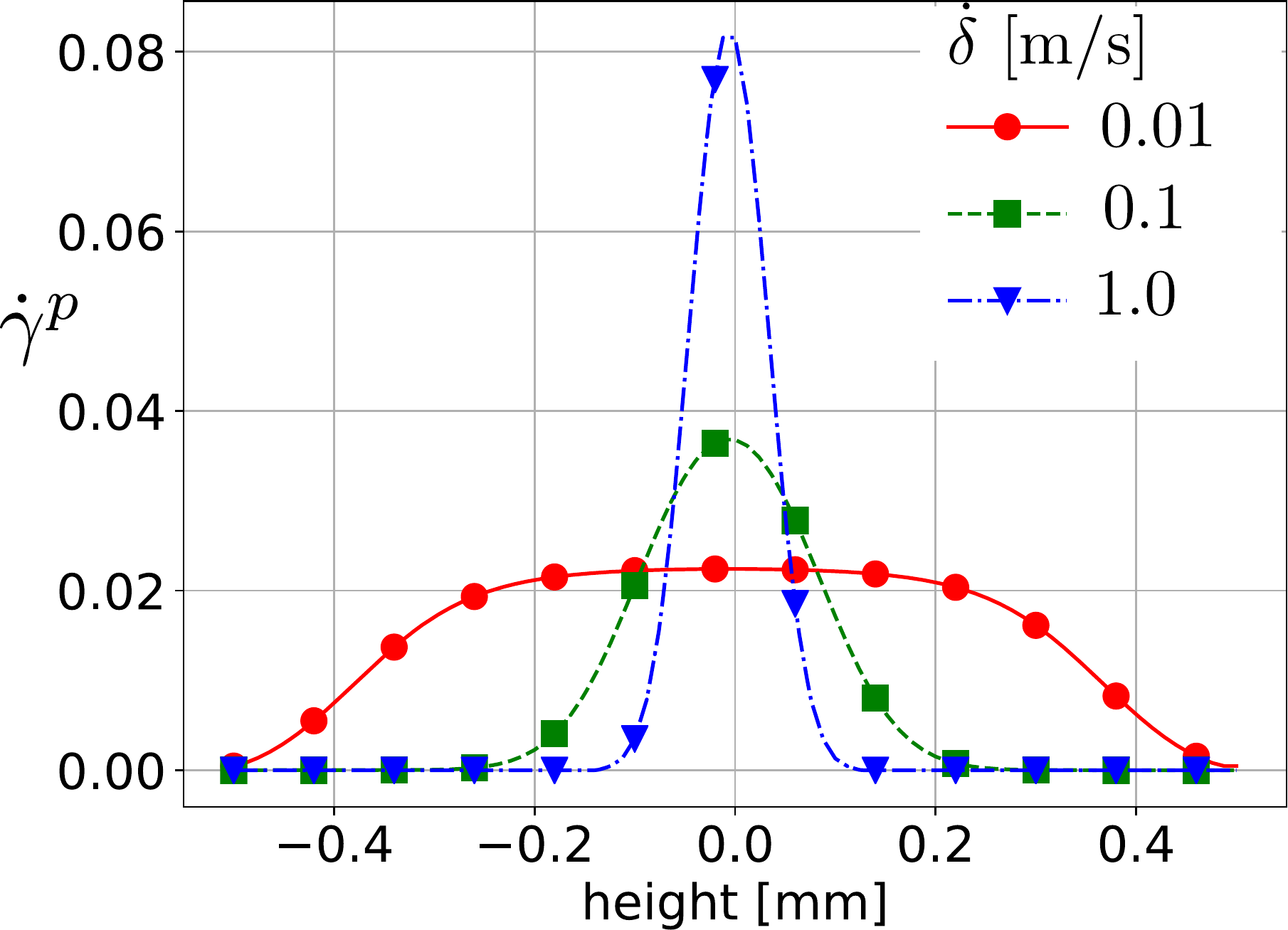}
  \end{minipage}%
  \caption{Left: $\tau-\delta$ response of the layer for different slip velocities $\dot{\delta}$ applied (velocity stepping). We observe that as the shearing rate increases, the softening behavior becomes more pronounced as a result of smaller localization widths due to the smaller characteristic diffusion time. Right: Profiles of strain localization rate inside the layer for different slip velocities $\dot{\delta}$ applied at the end of the analysis. Higher shearing velocities correspond to more localized plastic strain rate $\dot{\gamma}^p$ profiles.}
  \label{ch: 5 fig: tau_u_velocity_compare-l_loc_velocity_response}
\end{figure}

\begin{figure}[h!]
  \centering
  \begin{minipage}{.44\textwidth}
  \centering
  \includegraphics[width=0.9\linewidth]{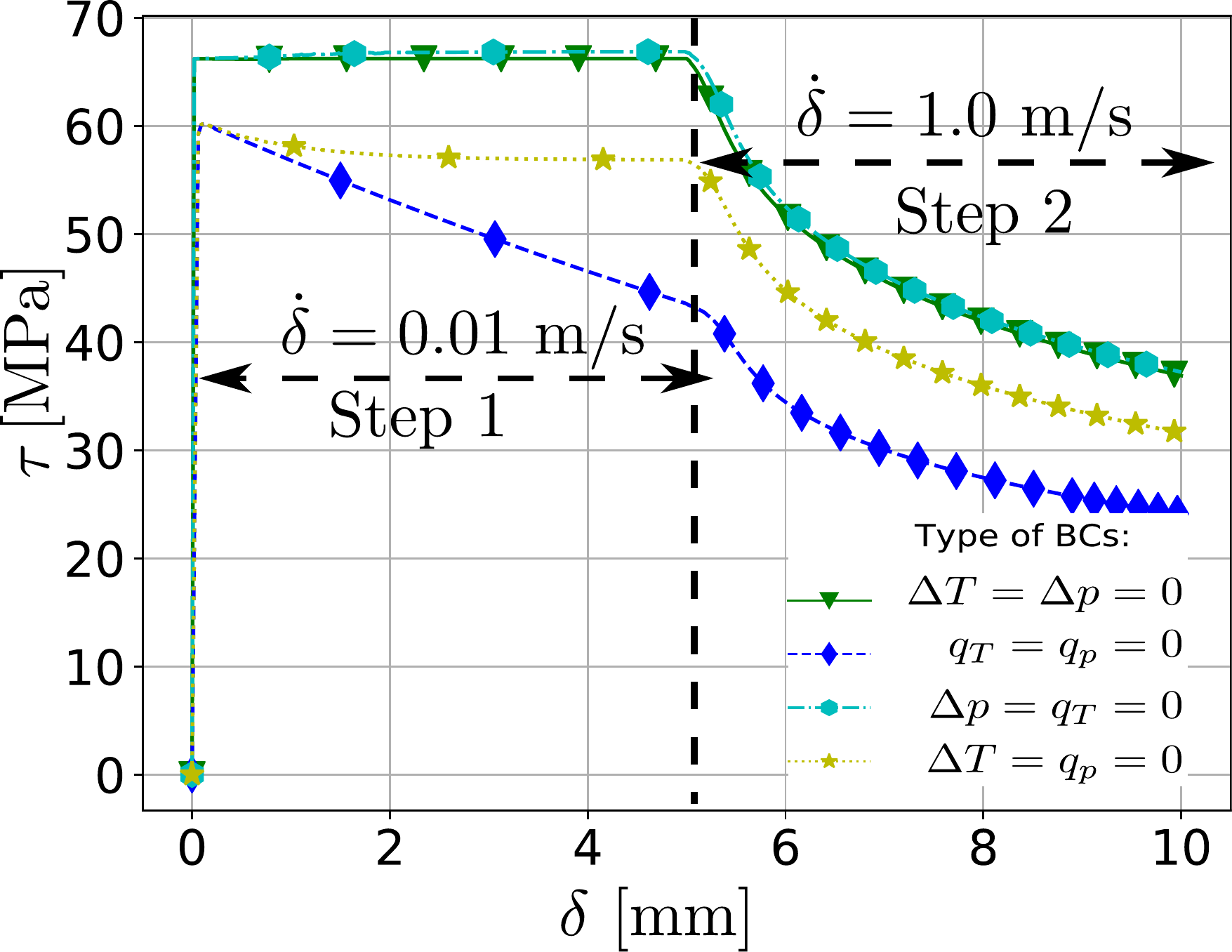}
  \end{minipage}\qquad
  \begin{minipage}{.46\textwidth}
  \centering
  \includegraphics[width=0.9\linewidth]{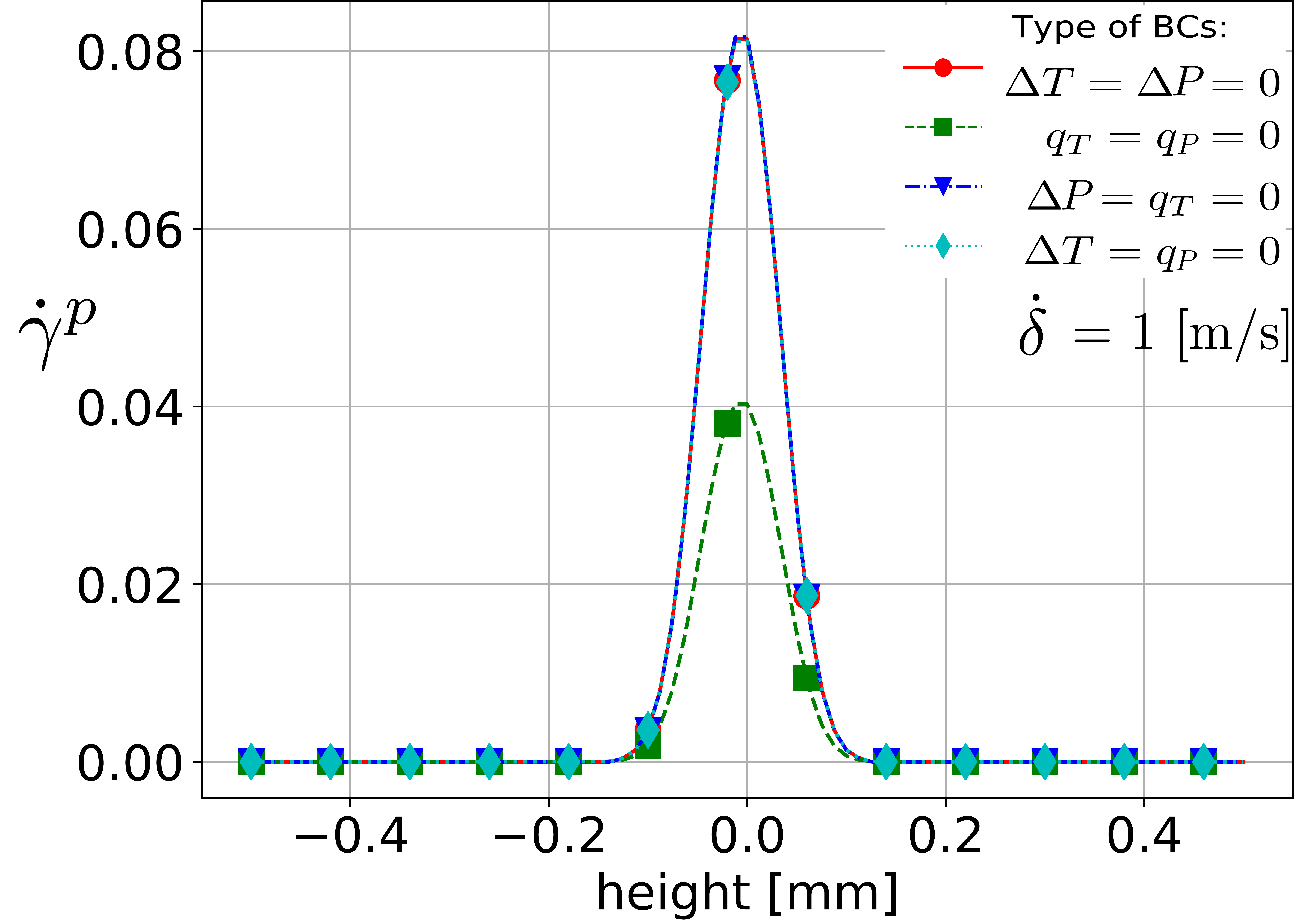}
  \end{minipage}%
  \caption{Left: $\tau-\delta$ response of the layer for different boundary conditions applied. An envelope is created between Isothermal drained ($\Delta T=\Delta P=0$) and Adiabatic-Undrained $(q_T=q_p=0)$ conditions. At the slow slip part of the analysis in the case of adiabatic undrained boundary conditions, thermal pressurization is present from the beginning. In this case, the initial stress at the start of the fast shear is lower and thus the stress drop is smaller. 
Right: Profiles of strain localization rate inside the layer for different boundary conditions. Since Cosserat material parameters and coseismic slip velocity $\dot{\delta}$ remain the same in all cases, the localization width does not change.}
  \label{ch: 5 fig: Tau_u_evolution_Loc_over_height_bcs_comparison}
\end{figure}

\noindent To better understand the effects of the applied shearing rate $\dot{\delta}$ in thermal pressurization and the overall effects of the boundary, as mentioned above, we proceed with the application of a velocity stepping shearing procedure. After consolidation (see Table \ref{ch: 5 table:three_step_procedure}), the layer is sheared with varying slip velocity $\dot{\delta}$ in two steps. At each step a target displacement $\delta$ of 5 mm at each end is reached for a total of 10 mm at the end of the analysis. The shear velocity $\dot{\delta}$ during the first shear step is $0.01$ m/s. For the second (final) step we ran different analyses under different applied constant shear velocity. The range of shear velocity values spans from $0.01$ m/s to $1$ m/s.
\begin{table}[h!]
\begin{center}
\begin{tabular}[]{l l l l l l}
\hline
\multicolumn{2}{l}{STEP} &Slip $\delta$ mm &\multicolumn{3}{c}{Slip velocity $\dot{\delta}$ m/s }\\
\hline
\hline
0 &Consolidation &- &\multicolumn{3}{c}{-}\\
1 &Shear &5 &\multicolumn{3}{c}{0.01}\\
2 &Shear &5 &0.01 &0.1 &1.0\\
\hline
\end{tabular}
\end{center}
\caption{Loading program for the analyses performed using the three step procedure.}
\label{ch: 5 table:three_step_procedure}
\end{table}\\ 
\newline
\noindent We intent to investigate the effect of the shearing rate on the frictional response of the layer. We investigate the predictions of our model concerning the apparent softening in the layer's frictional response subjected to isothermal ($\Delta T$=0) drained ($\Delta P$=0) boundary conditions. In Figure \ref{ch: 5 fig: tau_u_velocity_compare-l_loc_velocity_response} we compare the different shear stress $\tau$ seismic slip displacement $\delta$ responses for the different velocities applied at the final step of the analyses and we show the profiles of strain localization rate $\dot{\gamma}^p$ over the layer's height. We observe that the increase of slip velocity $\dot{\delta}$ has a weakening effect on the $\tau-\delta$ diagram as observed also by \cite{ Rattez2018b,Rattez2018a}. This happens due to the fact that a fast increase in the heat production term of equation \ref{ch: 5 Energy_balnca_eq} leads to an increase in the thermal pressurization term of equation \ref{ch: 5 mass_balance_main}, which in turn increases pressure and intensifies weakening due to the application of Terzaghi principle $\sigma^{eff}=\sigma_n+p$, ($p>0$ water pressure, $\sigma^{eff},\sigma_n>0$ in tension). The increase of slip velocity leads also to narrower localization zones, which are in agreement with the steeper post-peak response observed in $\tau,{\delta}$ diagrams.
\\
\noindent Finally, we investigate the influence of the boundary conditions of pressure and temperature to the behavior of the problem. Their influence to the frictional response is of great importance as they control the effect of diffusion on leading temperature and pressure increase ($\Delta T, \Delta P$) away from the localized zone. On the left part of Figure \ref{ch: 5 fig: Tau_u_evolution_Loc_over_height_bcs_comparison} we present the curves of $\tau-\delta$, for slip velocity $\dot{\delta}$ at the final step of the analysis of 1 m/s, for adiabatic-undrained ($q_T=q_p=0$), isothermal-drained ($\Delta T=\Delta P=0$), isothermal-undrained ($\Delta T=q_p=0$) and adiabatic-drained conditions ($q_T=\Delta P=0$). We observe that undrained boundary conditions influence the response the most as they tend to follow on the solution of uniform adiabatic undrained shear described in \cite{lachenbruch1980frictional} for small slip velocities $\dot{\delta}=0.01$ m/s. The difference at the peak strength between drained and undrained conditions has to do with the frequency our algorithm saves the output as well as the time increment used by the analysis (automatic time stepping). Furthermore, on the right of Figure \ref{ch: 5 fig: Tau_u_evolution_Loc_over_height_bcs_comparison} we present the plastic strain-rate profiles $\dot{\gamma}^p$ for different boundary conditions applied at the end of the analysis. We observe that for the given seismic slip of 10 mm, localization width is dependent on the seismic slip velocity applied and not on the boundary conditions.
\\
\newline
\noindent Based on the above results, we conclude that the true response of the fault gouge is very much dependent on the applied boundary conditions. Normally a kind of Robin boundary condition should be employed to better approximate the physical conditions. However, since this interaction between fluxes and essential boundary conditions is not yet sufficiently documented in the existing literature, the isothermal ($\Delta T =0$), drained ($\Delta p=0$) boundary conditions are closer to the real conditions due to the highly damaged regions encapsulating the fault gouge.\\
\newline
\noindent More specifically, the thermal diffusivity of the fault gouge and the surrounding material in the damaged zone (surrounding folliations) presents less variations than the corresponding hydraulic diffusivities. Thermal diffusivity of the fault gouge material can be one order of magnitude smaller than the diffusivity of the damaged zone \cite[see for instance][]{passelegue2014influence,tanaka2007thermal,yao2016crucial}. For the much more crucial hydraulic diffusivity parameter \cite[see][]{aydin2000fractures}, the hydraulic diffucivity ratio between the mature fault gouge and the surrounding folliated rock can be shown to differ up to 4 orders of magnitude, with diffusivity of the folliated rock being greater than that of the gouge. Furthermore, according to in situ observations \cite[see][]{ingebritsen2019earthquake} of increased water discharge to aquifiers, the ratio is expected to increase even more during coseismic slip. Therefore, we can estimate the ratios of thermal, $r_{th}$, and hydraulic, $r_{hy}$, diffusivities between the fault gouge material and the nearby damaged material:
\begin{align}
r_{th}=\frac{c^{rc}_{th}}{c^{fg}_{th}}\sim 10,\;\;\;r_{hy}=\frac{c^{rc}_{hy}}{c^{fg}_{hy}}\sim 10^4
\end{align} 
\noindent These parameters, further justify our choice of setting the boundary conditions in the rest of the paper to isothermal $\Delta T=0$ drained $\Delta P=0$. 
\subsection{Shearing of a mature fault under moderate slip ($\delta=100$ mm), and variable slip velocities}\label{ch: 5 sec: Two_step_procedure}
\begin{figure}[h!]
\centering
  \includegraphics[width=0.9\linewidth]{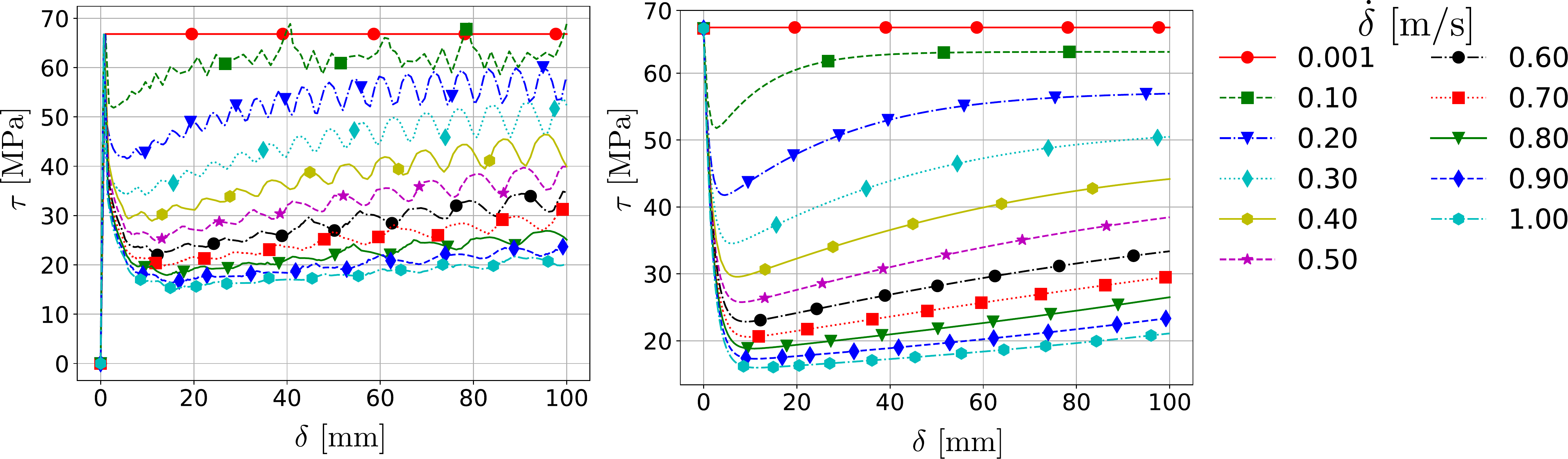}
  \caption{Left: $\tau-\delta$ response of the layer for different velocities. Isothermal, drained boundary conditions $\Delta T=\Delta P=0$ are applied. 
  Frictional strength regain is observed due to the diffusion at the boundaries. The frictional response presents oscillations due to the traveling plastic strain rate instability. A smaller residual friction value is achieved. 
 Right: Fitted $\tau-\delta$ response of the layer for different velocities, by filtering out the oscillations resulting form traveling instabilities. Isothermal, drained boundary conditions $\Delta T=\Delta P=0$ are applied. All analyses at the start reach the peak strength $\tau =66.67\;$MPa. As slip $\delta$ progresses an increase of the residual shear stress $\tau$ takes place. }
  \label{ch: 5 fig: tau_u_velocity_compare-fit}
\end{figure}

\begin{figure}[h!]
  \centering
  \advance\leftskip-4cm
  \advance\rightskip-4cm
  \begin{minipage}{.55\textwidth}
  \centering
  \centerline{\includegraphics[width=0.9\linewidth]{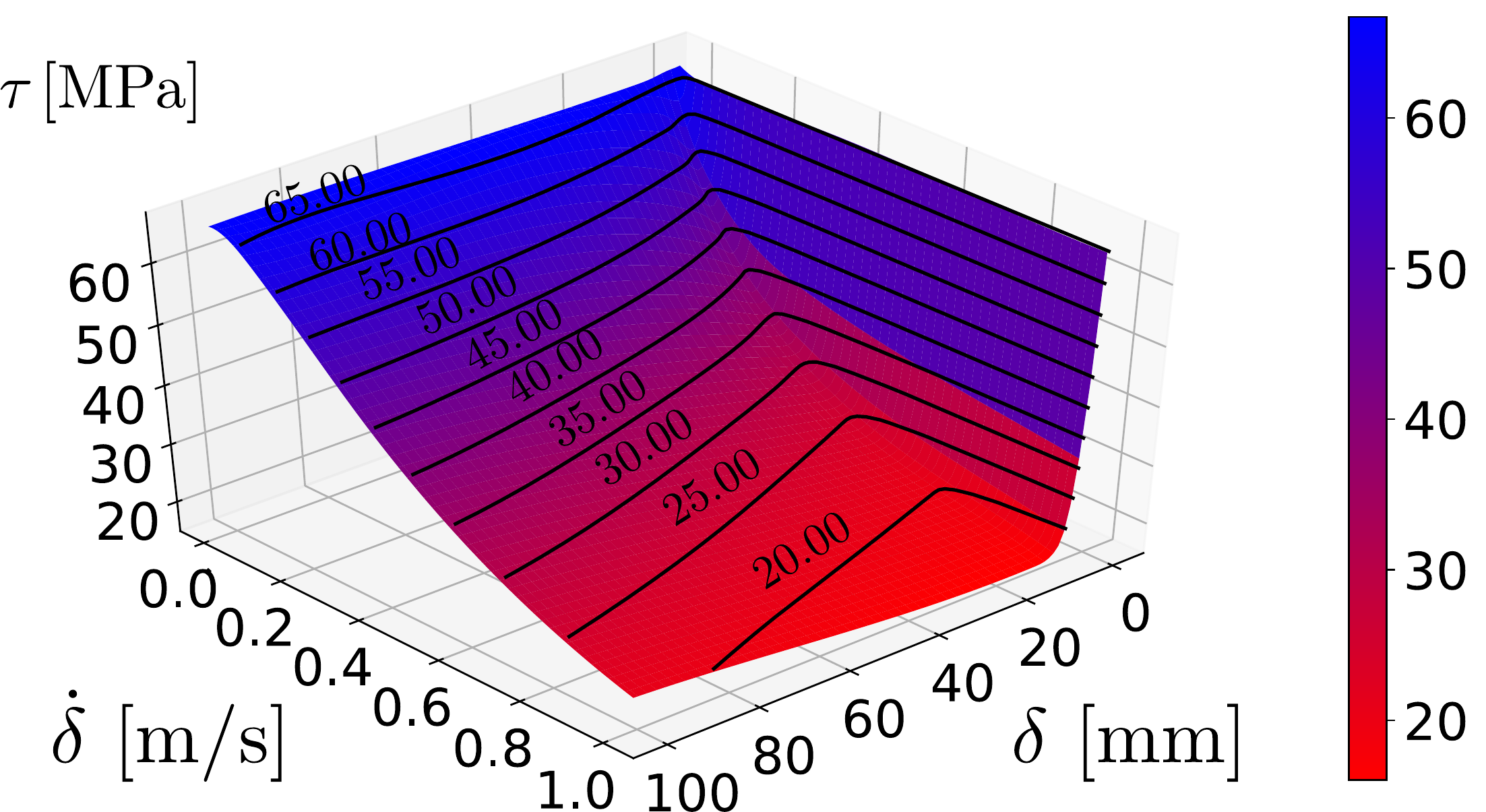}}
  \end{minipage}\qquad
  \begin{minipage}{.55\textwidth}
  \centering
  \centerline{\includegraphics[width=0.9\linewidth]{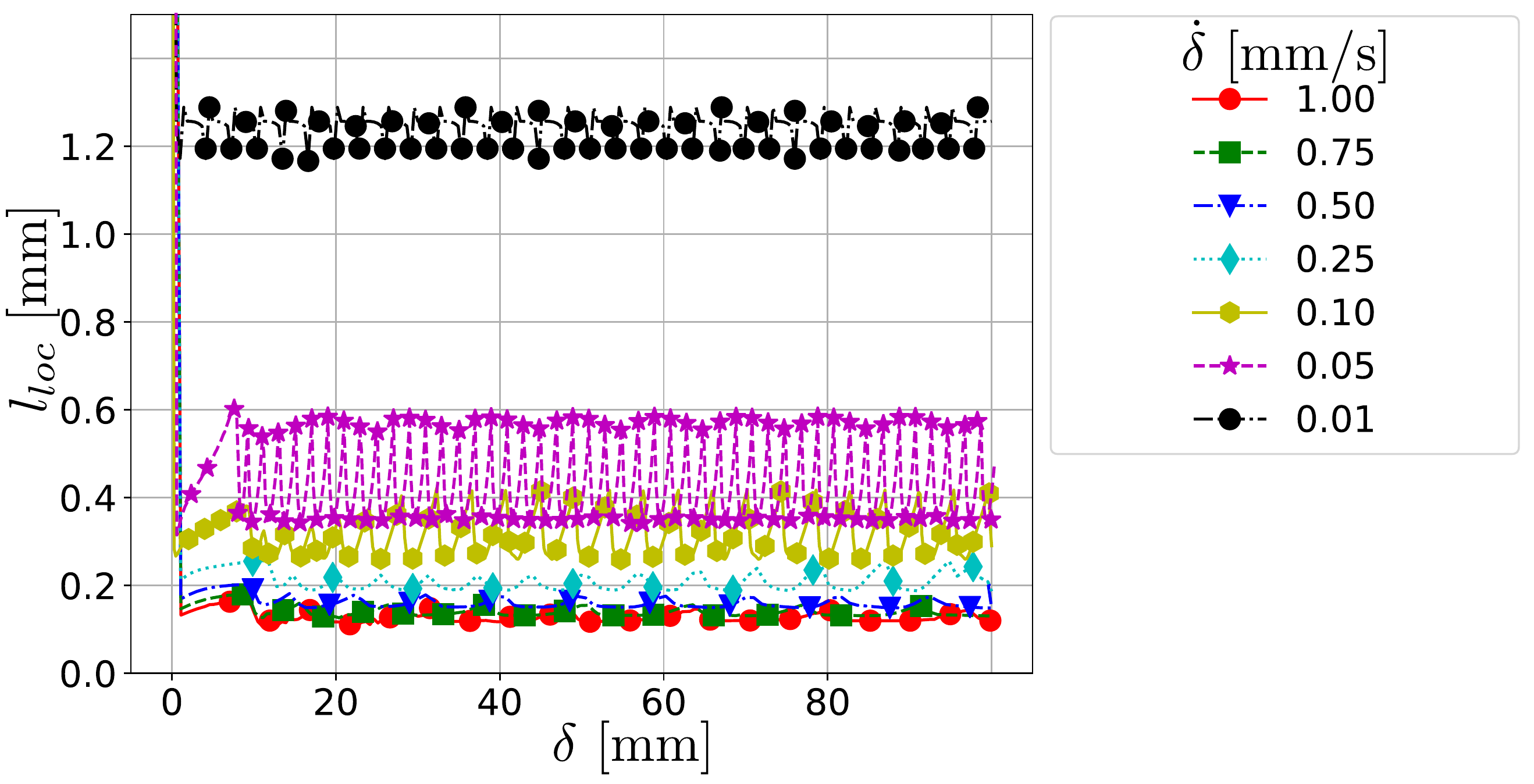}}
  \end{minipage}%
  \caption{Left: 3D fitted surface of $\tau$ with slip distance $\delta$ and velocity $\dot{\delta}$. Right: $l_{loc}-\delta$ response of the localization width inside the layer for different boundary shear velocities applied at the boundaries.
  We notice that the localization width is oscillating for the small to intermediate range of shear velocities $\dot{\delta}=0.01-0.25$ m/s. This is due to the interaction between the diffusion lengths of pressure and temperature. }
  \label{ch: 5 fig: tau_u_velocity_compare-frictional-surface}
\end{figure}

\noindent 
To better illustrate the dependence of the fault behavior to the velocity of seismic slip $\dot{\delta}$, we run a second part of analyses for the case of isothermal drained conditions ($\Delta T=\Delta P=0$) in which the intermediate part of slow shear velocity has been omitted and the fault model is immediately subjected to fast slip velocity rates after initial consolidation. Furthermore, the target seismic displacement $\delta$ has been increased to $100.0$ mm. We aim that way to examine in more detail the fault's response under displacement scales commonly observed in nature at variable slip rates. Fast slip rates would correspond to the fault gouge being in the center of the fault rupture area. For the numerical steps of the simulations see Table \ref{ch: 5 table:two_step_procedure}.
\begin{table}[h!]
\begin{center}
\begin{tabular}[]{l l l l l }
\hline
\multicolumn{2}{l}{ STEP } &Height $\text{H}$ mm &Slip $\delta$ mm &Slip velocity $\dot{\delta}$ m/s  \\
\hline
\hline
0 &Consolidation &1 &- &-\\
1 &Shear & &100.0 &$\{0.01,0.05,0.1,0.25, 0.50,0.75,0.90,1.0\}$\\
\hline
0 &Consolidation &1 &- &-\\
1 &Shear & &1000.0 &1.0\\
\hline
0 &Consolidation &1 &- &-\\
1 &Shear & &1000.0 &0.5\\
\hline
0 &Consolidation &2 &- &-\\
1 &Shear & &1000.0 &0.5\\
\hline
\end{tabular}
\caption{Loading program for the analyses performed using the two step procedure.}
\label{ch: 5 table:two_step_procedure}
\end{center}
\end{table}\\
\noindent In Figure \ref{ch: 5 fig: tau_u_velocity_compare-fit}, we present the frictional response of the fault gouge, shear stress $\tau$, with the seismic slip $\delta$ on top of the layer. We plot the $\tau-\delta$ response for different values of seismic slip velocity. It can be clearly seen from the results that two behaviors are present depending on the shear velocity. If the slip velocity is low, then the layer accommodates the heat produced from the plastic work during yielding of the material and both heat and pressure diffuse efficiently away from the yielding zone, which has a comparatively large localization width $l_{loc}$ as shown in Figure \ref{ch: 5 fig: tau_u_velocity_compare-frictional-surface}. As slip velocity increases the post peak softening response is seen in larger parts of the analysis before eventually diffusion dominates and peak shear strength is restored. However, shear strength is only partially restored for the analyses of large shear velocities.
\\
\newline
\noindent In the left part of Figure \ref{ch: 5 fig: tau_u_velocity_compare-fit} the frictional response ($\tau - \delta$) of the layer obtained from the numerical analyses of first row of Table \ref{ch: 5 table:two_step_procedure} is presented. We notice the existence of oscillations in the frictional response for all velocities apart from the very small ($\dot{\delta}=0.01$ m/s). The oscillations start during the apparent softening branch of the analysis frictional response. They are affected by the boundaries of the model, namely in the case of undrained, adiabatic boundaries the shear band travels to one of the boundaries and then persists at this position, while for isothermal drained conditions an oscillatory response is present (see also sections \ref{ch: 5 Traveling instability},\ref{ch: 5 Traveling_instability_new} for more details about this important finding).\\
\newline
In the right part of Figure \ref{ch: 5 fig: tau_u_velocity_compare-fit} we present the $\tau-\delta$ fit of the numerical results. We employ a function composed of an exponential decay and a logistic curve.
\begin{align}
\tau(\delta,\dot{\delta})=a(\dot{\delta}) \exp{\left[-b(\dot{\delta}) \delta\right]} +\frac{c(\dot{\delta})}{d(\dot{\delta})+\exp\left[\right.-f(\dot{\delta})(\delta-\left.g(\dot{\delta}))\right]},
\end{align}
\noindent where $\alpha(\dot{\delta}),\;b(\dot{\delta}),\;c(\dot{\delta}),\;d(\dot{\delta})$ are the interpolation parameters, dependent on $\dot{\delta}$. The fit is used to simplify conceptually the results and highlight the main findings of the numerical analyses. It contains the initial frictional weakening and subsequent frictional regain due to the diffusion at the boundaries of the model. Furthermore, the effect of the oscillations is highlighted. The fit passes through the average of the oscillations of the numerical analyses, indicating that due to the oscillations, friction is not fully restored to its initial value. There exists a residual value of friction at the later stages of the slip. \\
\newline
\\
\noindent On the left part of Figure \ref{ch: 5 fig: tau_u_velocity_compare-frictional-surface}, we present the frictional surface that corresponds to the fit of the previous paragraph. Through the use of two dimensional interpolation of the results of Figure \ref{ch: 5 fig: tau_u_velocity_compare-fit}, we are able to estimate the frictional response of the fault gouge over a region of low to moderate seismic slips ($\delta=0\sim0.1$ m) and seismic slip velocities ($\dot{\delta}=0.1\sim 1$ m/s). 
\\
\newline
\noindent On the right part of Figure \ref{ch: 5 fig: tau_u_velocity_compare-frictional-surface}, we present the evolution of the shear band width for the different seismic slip velocities. In order to estimate the localization width in each case a curve according to equation \eqref{ch: 5 l_dot_curve_fit_Rice} described in \cite{rice2006heating} was selected for fitting.
\begin{align}
\dot{\gamma}^p = &A+\frac{B}{\sqrt{1\pi}D}\exp\left[-\frac{1}{2}{\left(\frac{y-C}{D}\right)}^2\right]\nonumber\\
l_{loc} = &2\sqrt{2\ln(2)}D
\label{ch: 5 l_dot_curve_fit_Rice}
\end{align}
\noindent It is clear that large velocities lead to narrower localization widths $l_{loc}$. We observe that for large velocities localization width is not monotonously decreasing, but rather it exhibits some noise as shearing progresses. This goes beyond the results of \cite{Rattez2018a,Rattez2018b}, where the localization width was shown to progressively decrease until it remains constant. This behavior has to do with the fact that the instability exhibited here by the material is a traveling wave instability.\\
\newline
\noindent This jerky behavior of frictional oscillations, which reminds the Portevin Le Chatelier effect can be responsible for higher frequency instabilities during seismic slip and enhance the frequency content of an earthquake event as discussed in \cite{Aki1967a, BRUNEJN1970, Haskell1964, Tsai2020}. The observed behavior of frictional regain and frictional oscillations is primarily due to the applied isothermal drained boundary conditions that allow for a traveling PSZ inside the fault gouge. In the case of adiabatic undrained conditions the shear band can be shown to travel towards a boundary, where it is trapped for the duration of the analysis and the results obtained in this case are closer to those derived in the case of uniform shear (\cite{lachenbruch1980frictional}).
\subsection{Shearing of a mature fault for large slip $\delta=1\;\text{m}$ and coseismic slip velocity $\dot{\delta}=1\;\text{m/s}$. \label{ch: 5 Traveling instability}}
\begin{figure}[h!]
\centering
\includegraphics[width=0.45\linewidth]{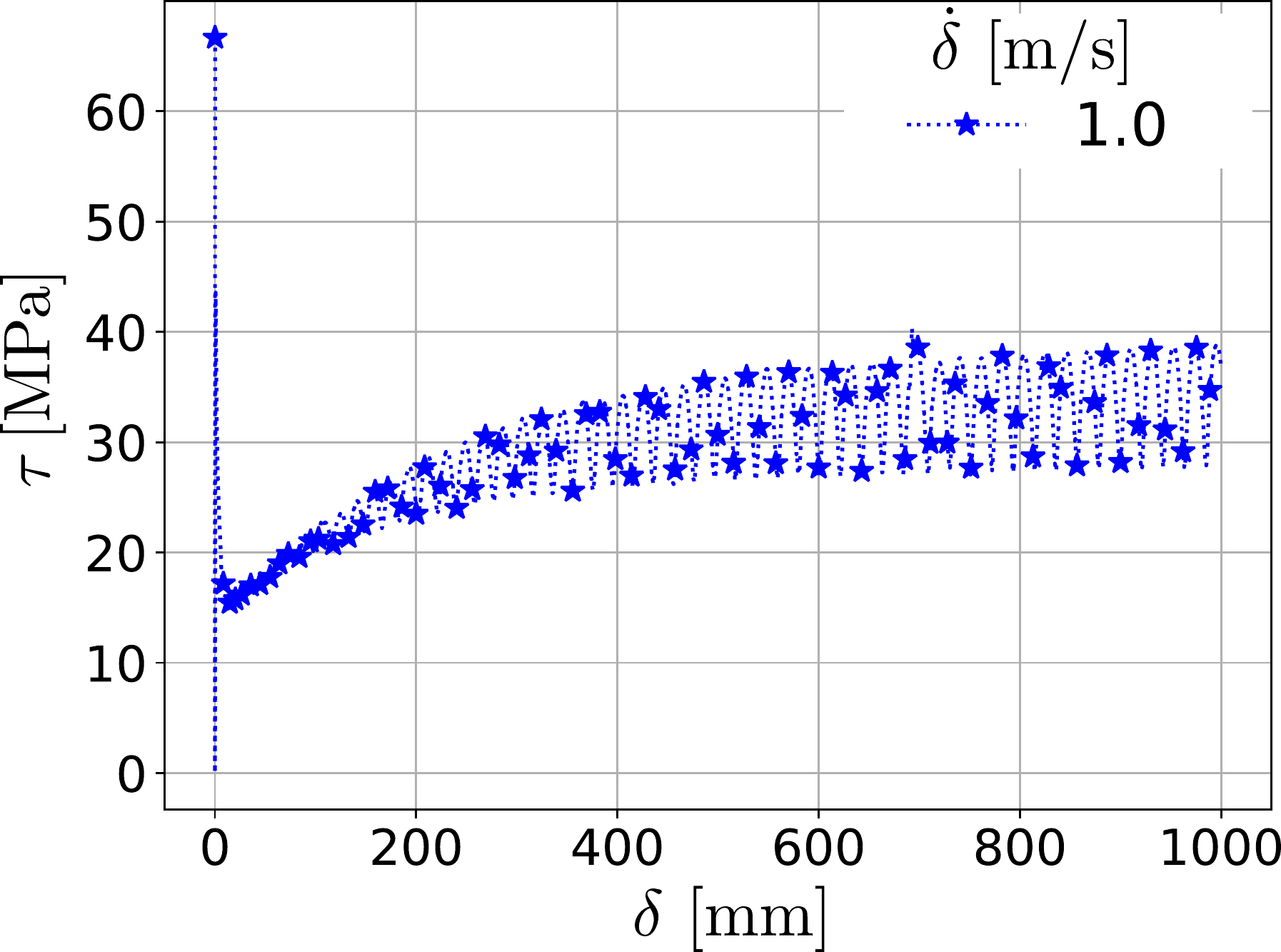}
\caption{ Left: Evolution of $\tau$ with slip distance $\delta$. We observe that after sufficient time has passed the oscillations have stabilized in amplitude and frequency partially recovering the layer's initial shear strength. The steady state reached around which the oscillations take place is reached after a slip of 1.0 m. The value of friction at the steady state is a result of the interplay between the rate of work dissipated into heat and the diffusion properties of the fault gouge.\label{ch: 5 fig: tau_u_velocity_compare_1}}
\end{figure}
\begin{figure}[h!]
  \centering
\advance\leftskip-1cm
 \begin{minipage}{.55\textwidth}
  \centering
  \centerline{\includegraphics[width=0.9\linewidth]{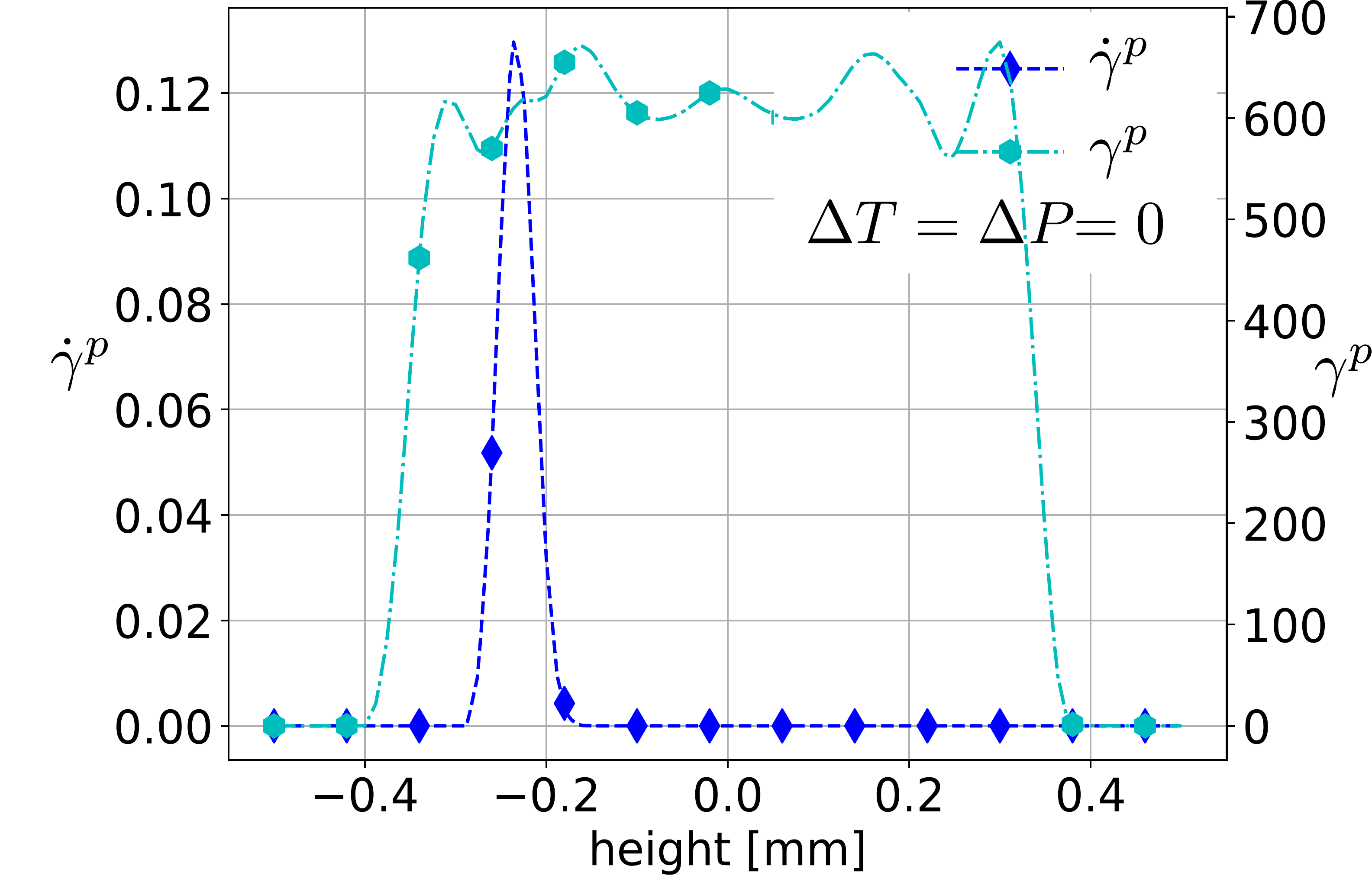}}
  \end{minipage}
  \begin{minipage}{.55\textwidth}
  \centering
  \centerline{\includegraphics[width=0.9\linewidth]{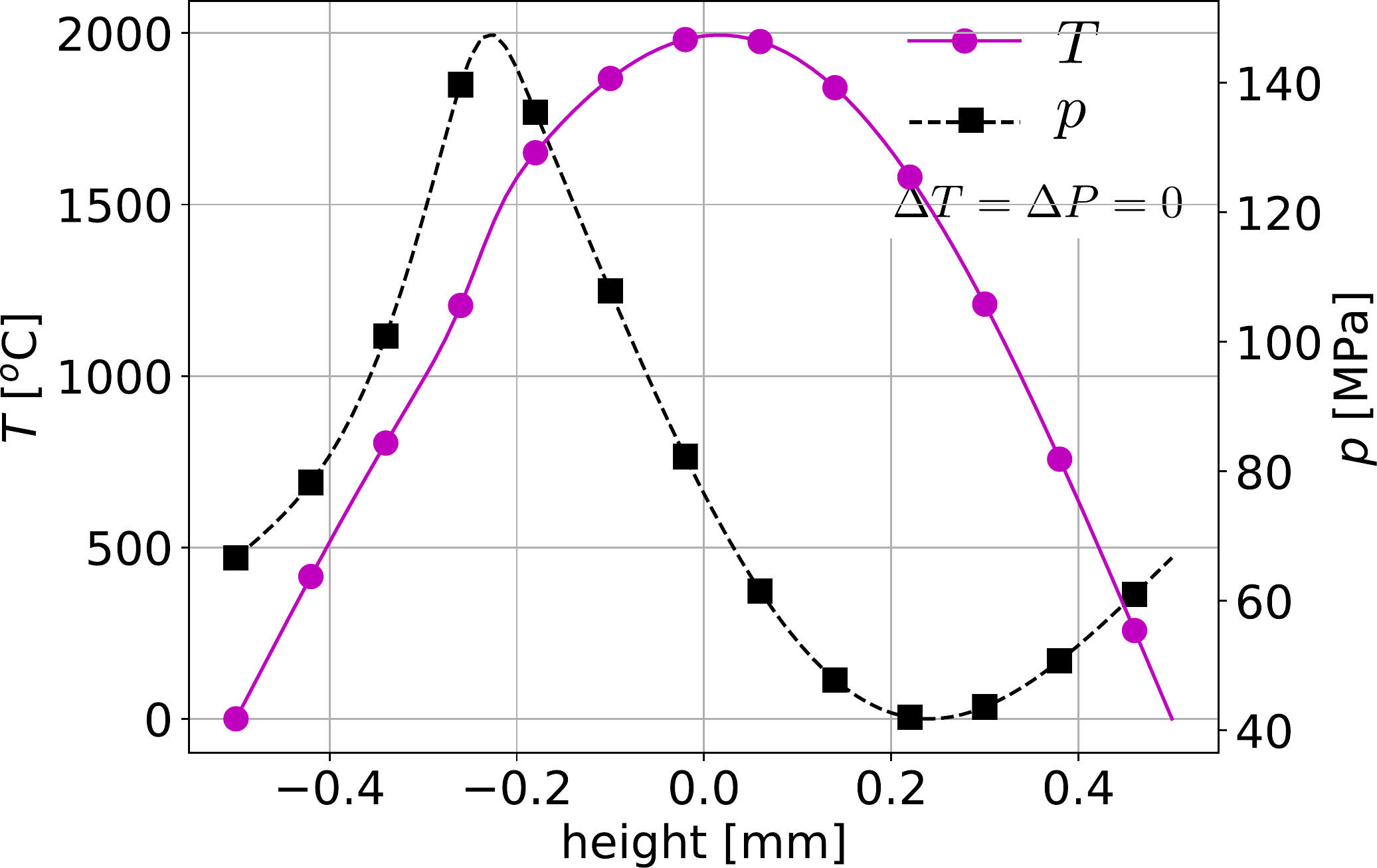}}
  \end{minipage}
  \caption{Left: Profiles of shear strain rate and accumulated plastic shear strain $\dot{\gamma}^p,\gamma^p$ at the end of the analysis for applied slip $\delta=1$ m and slip rate $\dot{\delta}=$1 m/s. Since the two profiles differ, we conclude that the localization oscillates inside the layer. Localization does not travel the whole of the layer due to the boundary conditions applied. Right: Profiles of pressure and temperature $p,\; T$ at the end of the analysis for applied slip $\delta=1$ m and slip rate $\dot{\delta}=$1 m/s. Diffusion at the boundaries leads to extremely high values of temperature $\ T=$2000$\;^{o}$C.}
   \label{ch: 5 fig: l_dot_gamma_final_1000-T_p_final_1000}
\end{figure}

\noindent In order for the observed oscillations to fully develop in amplitude for the analyses with high slip rate $\dot{\delta}$ we apply a very large shear displacement. Figure \ref{ch: 5 fig: tau_u_velocity_compare_1}, presents the $\tau,\delta$ response for a slip velocity $\dot{\delta}$ of 1 m/s and an applied slip $\delta=1$ m. As can be seen from the above analysis, the shear strength of the layer is eventually oscillating around a new residual strength value, which is smaller than the original peak strength.
\\
\newline 
\noindent The left part of Figure \ref{ch: 5 fig: l_dot_gamma_final_1000-T_p_final_1000} shows the profiles of plastic strain rate $\dot{\gamma}^p$ and accumulated plastic strain $\gamma^p$ at the end of the coseismic slip. It is clear that the shear band travels across the fault gouge layer, since the accumulated plastic strain profile is larger in width than the localization width of the instability. This is one major difference compared to small slip rates, which our analyses under large displacements allowed to highlight (see Figure \ref{ch: 5 fig: tau_u_velocity_compare-fit}). Finally, the right part of Figure \ref{ch: 5 fig: l_dot_gamma_final_1000-T_p_final_1000} presents the profiles of temperature $T$ and pressure $p$ at the end of the analysis. We observe that the temperature reached is much higher than the one required for the onset of melting for the minerals present in the seismogenic zone, see \cite{rice2006heating}. This has to do with the relatively high friction coefficient $\mu$ used in our analyses. A moderate value of $\mu$=0.25 would roughly halve the temperature observed. This does not preclude though other frictional weakening mechanisms, such as thermal decomposition of minerals (see \cite{Sulem2009,Sulem2016,Veveakis2014,Alevizos2014a}), that might become dominant after thermal pressurization becomes impossible.
\\
\newline
\noindent A cycle of friction during the oscillation of the traveling shear band (PSZ) inside the fault gouge is separated in two stages: First the band travels inside the medium which corresponds to a weakening phase of the frictional response. The weakening phase takes place as the band travels across a hot region of the layer in which case according to equation \eqref{ch: 5 linearized_system} the pressure increases as $\Delta T$ is positive. Next, as the band travels inside the fault gouge expanding the yield zone, it approaches the boundaries. Close to the boundaries, temperature and pressure diffusion are more efficient. In particular, near the boundary, temperatures are lower as dictated by the parabolic profile of temperature $T$ (see Figure \ref{ch: 5 fig: l_dot_gamma_final_1000-T_p_final_1000}) due to diffusion. The high diffusion gradients result in the cooling of the region, where the shear band is present. This in turn leads to a decrease in the applied pressure, therefore, the layer regains part of its strength. During each cycle the yield region is slightly increasing and the oscillations grow in period, since the yield zone is progressively expanding, and in amplitude, since the cooling effect becomes more pronounced and the temperature gradients become steeper.
\subsection{Effect of the layer's height on the oscillations behavior \label{ch: 5 Traveling_instability_new}}
\begin{figure}[h!]
  \centering
  \advance\leftskip-4cm
  \advance\rightskip-4cm
  \begin{minipage}{.58\textwidth}
  \centering
  \includegraphics[width=0.9\linewidth]{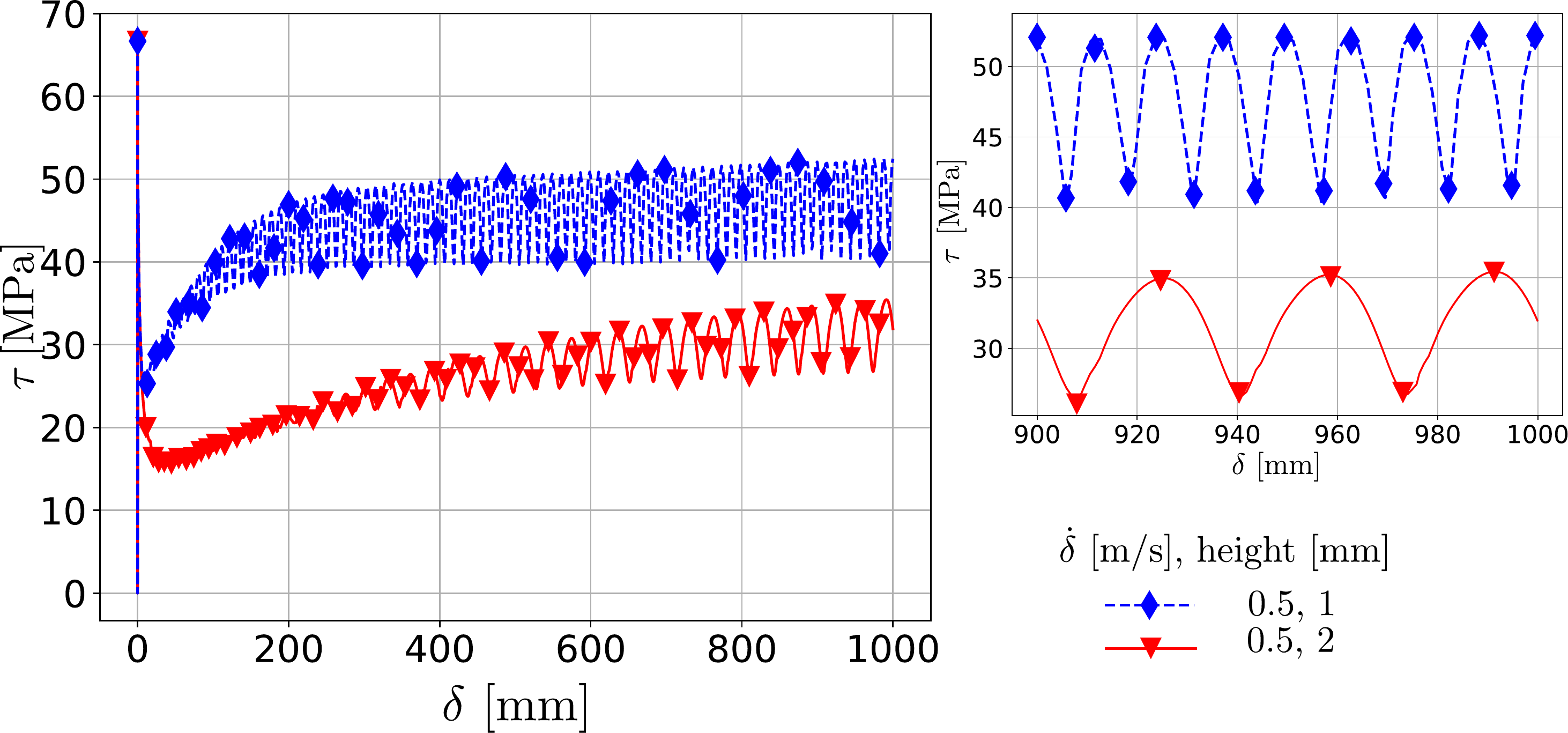}
  \end{minipage}\qquad
  \begin{minipage}{.42\textwidth}
  \centering
  \includegraphics[width=0.9\linewidth]{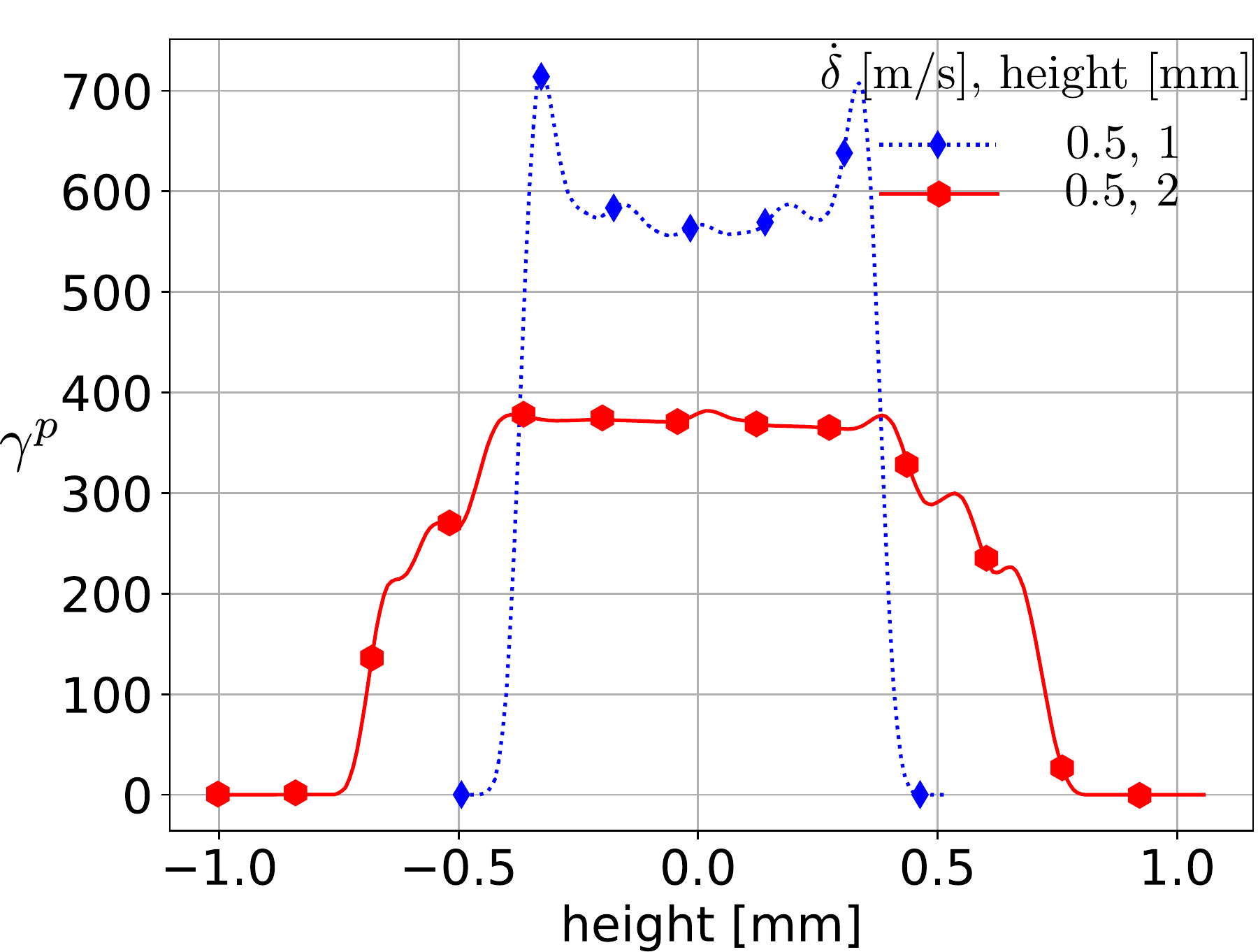}
  \end{minipage}%
  \caption{Left: Frictional strength evolution of two layers of width 1 mm (dashed , blue, diamond curve) and 2 mm (solid, red, triangle curve) under the same seismic slip velocity $\dot{\delta}= 0.5$ m/s. Right: Profiles of accumulated plastic strain rate at the end of the two analyses. Yielding has not yet fully developed in the case of thickness of 2 mm.}
  \label{ch: 5 fig: h_comparizon}
\end{figure}
\noindent In order to investigate how the boundaries affect the evolution of the traveling instability and of the frictional behavior inside the medium we compare the response between two layers of different height, 1 mm and 2 mm respectively, under a constant seismic slip velocity of 0.5 m/s (see Figure \ref{ch: 5 fig: h_comparizon}). We notice that the two layers exhibit a weakening response during the initial stages of thermal pressurization, however, the layer of thickness 2 mm reaches an overall lower minimum and a lower value for the residual shear strength. These values are essentially controlled by the diffusion processes. In the case of 1 mm, diffusion to the boundaries is more efficient and the layer regains more of its strength. Furthermore, diffusion is affected by the traveling instability inside the medium. In the case of the wider layer, from the period of the frictional strength oscillations (see inset of left part of Figure \ref{ch: 5 fig: h_comparizon}) we deduce that it takes almost twice the time for the traveling instability to cross through the layer. Therefore, temperature and pressure diffuse slower over the height of the layer. The period of the oscillations depends mainly on the height of the fault gouge layer.\\
\newline
\noindent The oscillations in the frictional response of the layer are dependent both in period and in amplitude on the height of the layer. Oscillations of higher amplitude occur in the case of the shorter layer height, where the plastic zone has more time to develop and the diffusion gradients close to the boundary are steeper. It is of paramount importance to take into account the slopes of $\tau$ w.r.t. $\delta$ both at the beginning and at the oscillation phase since they are vital for the instability nucleation in the various stick and slip models (see \cite{dieterich1992earthquake,ruina1983slip,Rice1973b}). For high velocities ($0.5\sim 1$ m/s) the slope at the beginning is the steepest and controls the energy balance (and the radiated energy), in contrast, for relatively small velocities ($0.1\sim 0.3$ m/s, see Figure \ref{ch: 5 fig: tau_u_velocity_compare-fit}) the trend might be different leading to radically different instability conditions. By calculating the radiated energy from an earthquake and having an accurate value for the seismic slip velocity $\dot{\delta}$, we can estimate the width of the fault gouge during the earthquake. \\
\newline
\noindent We note here that based on the discussion of section \ref{ch: 5 Normalized_system_of_equations}, applying a scaling that takes both into account the effect of coseismic velocity $\dot{\delta}$ and the characteristic height of the layer $\text{H}_0$, allows us to verify the above results. In essence we note that keeping the velocity constant, an increase in the layer's height affects only the diffusion terms. In particular, the characteristic diffusion time $t_0$ doubles in the case of the thicker layer, indicating that the frictional regain due to the boundary effects will take more to develop. Note also that doubling the height of the layer leads to the doubling of the period in the thicker layer, indicating that the frictional oscillations depend on the diffusion properties of the fault gouge and its boundaries, and not on the characteristic length of the microstructure (Cosserat radius $R$).  
\subsection{Proposed mechanism for a traveling PSZ}
\noindent We will proceed now to propose a mechanism that is responsible for the traveling of the strain localization (PSZ) inside the fault gouge. We consider initially two points inside the yielding region (PSZ): The first is lying in the middle of the localization while the second point lies in vicinity of the first one. Initially the temperature and the pore fluid pressure, increase faster in the middle of the layer. However, as the pore fluid pressure increases, friction inside the fault gouge drops due to the Terzaghi principle (apparent softening). This affects the thermal load due to dissipation inside the yielding region. In particular, the temperature increase in the middle is less than the corresponding temperature increase at the neighboring points of the yielding region, where dissipation is higher. Since temperature increases faster in the points near the middle than in the middle itself, the pore fluid pressure increase due to thermal pressurization is also greater in the neighborhood than in the middle.\\
\newline
\noindent The increase in the rate of pore fluid pressure indicates an increase in plastic strain rate in the neighboring points, because of the Terzaghi principle, which holds true also for the rates of fault friction and pore fluid pressure ($\dot{\tau}(t)=-\dot{p}(t)$). This increase of the plastic strain rate in the neighboring points corresponds to yielding accumulating away from the middle point. Thus the yielding region (i.e. the PSZ) starts traveling, following the accumulation of plastic strain rate. This leads to oscillations in the frictional response of the fault gouge. The crests of the oscillations correspond to the position of the PSZ lying closer to the boundaries of the fault gouge, while the troughs correspond to the PSZ in the middle of the fault gouge layer. The above mechanism embodies the finiteness of the yielding region. 
\section{Introduction of viscosity - rate and state phenomenology \label{ch: 5 viscosity_reference}}
\begin{figure}[h]
  \centering
  \includegraphics[width=0.5\linewidth]{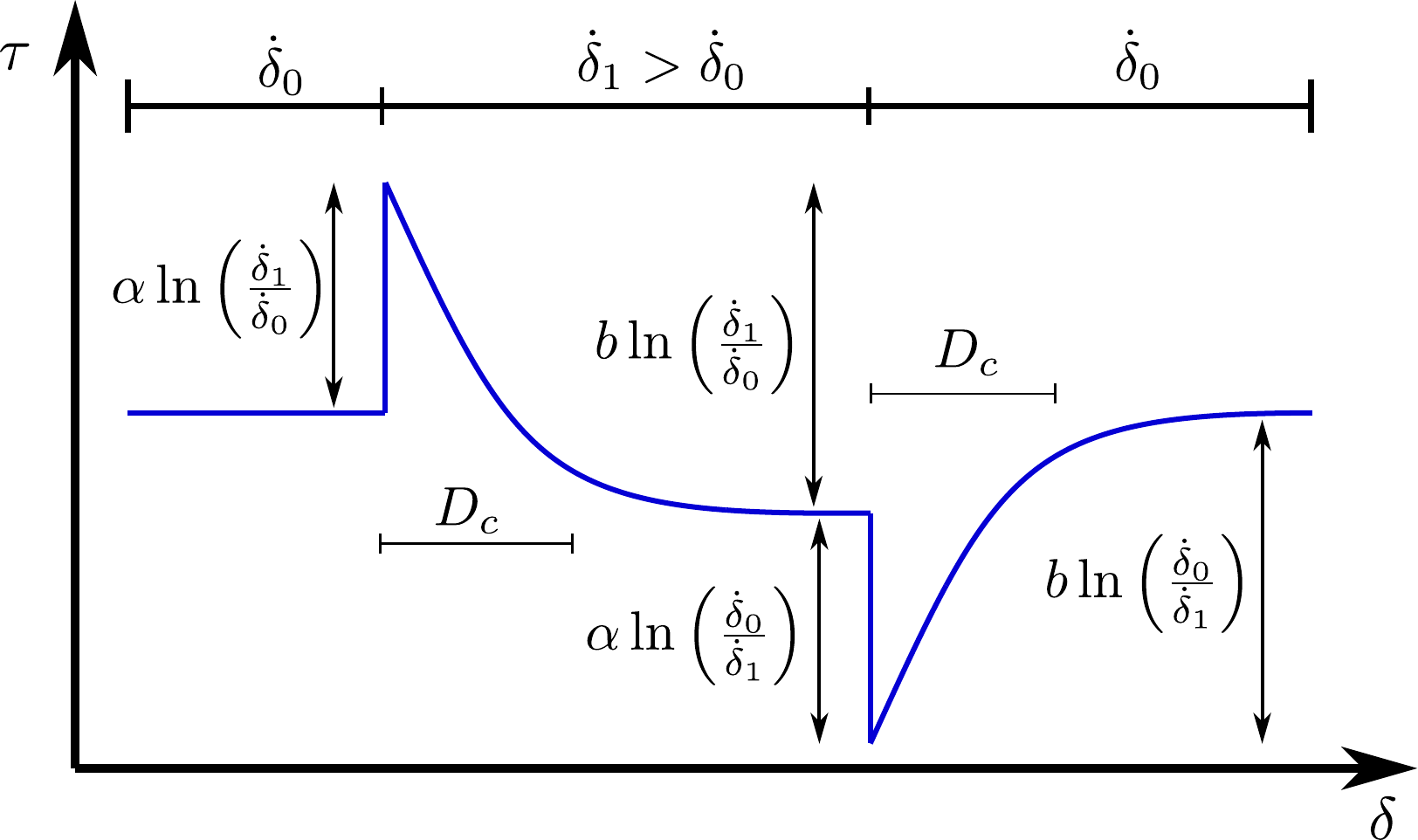}
  \caption{Rate and state phenomenology \cite[image taken from][]{tzortzopoulos2021absorbent}.}
  \label{ch: 5 fig: rate and state}
\end{figure}

\begin{figure}[h!]
  \centering
  \begin{minipage}{.45\textwidth}
  \centering
  \includegraphics[width=0.9\linewidth]{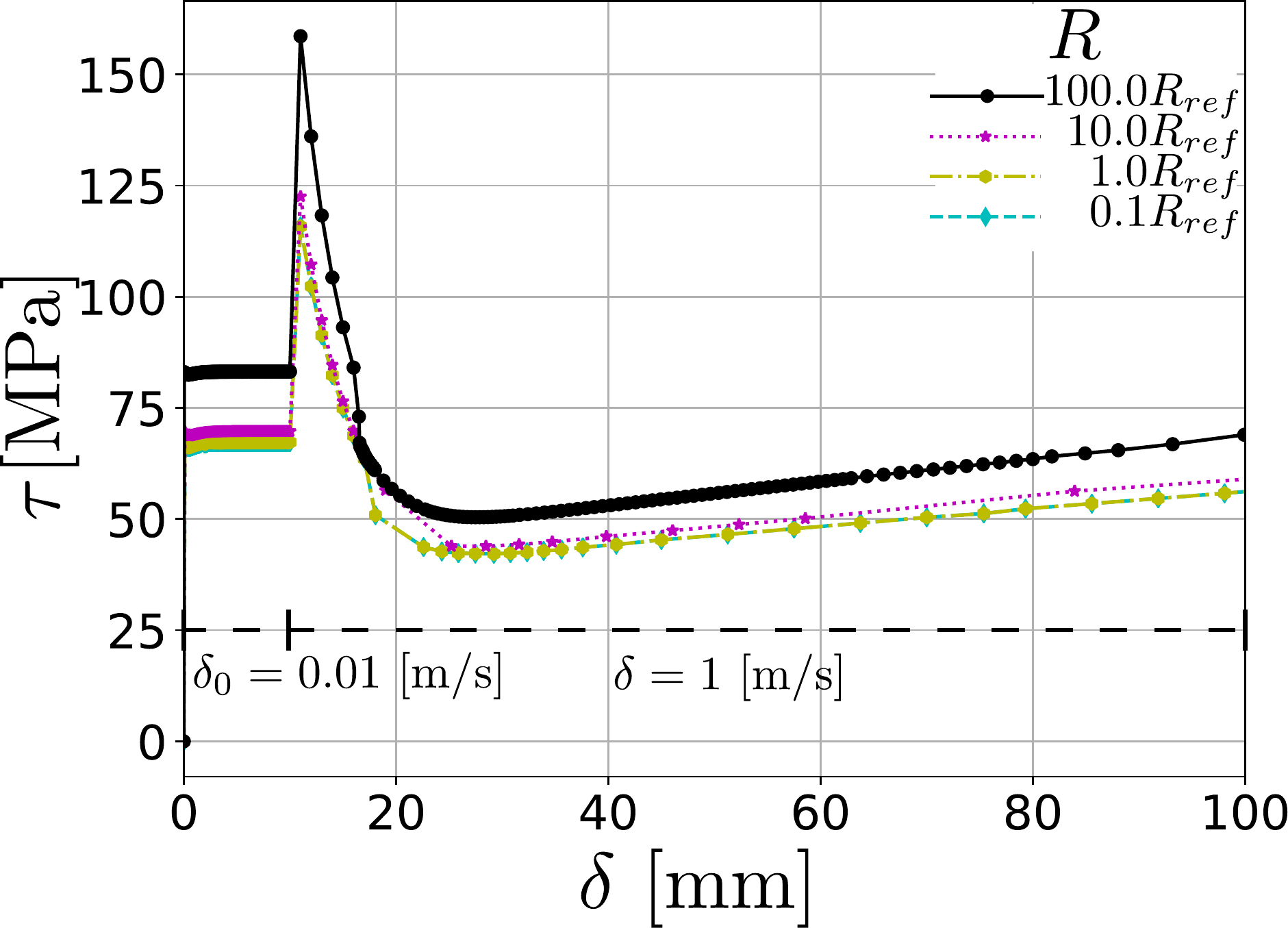}
  \end{minipage}\qquad
  \begin{minipage}{.45\textwidth}
  \centering
  \includegraphics[width=0.9\linewidth]{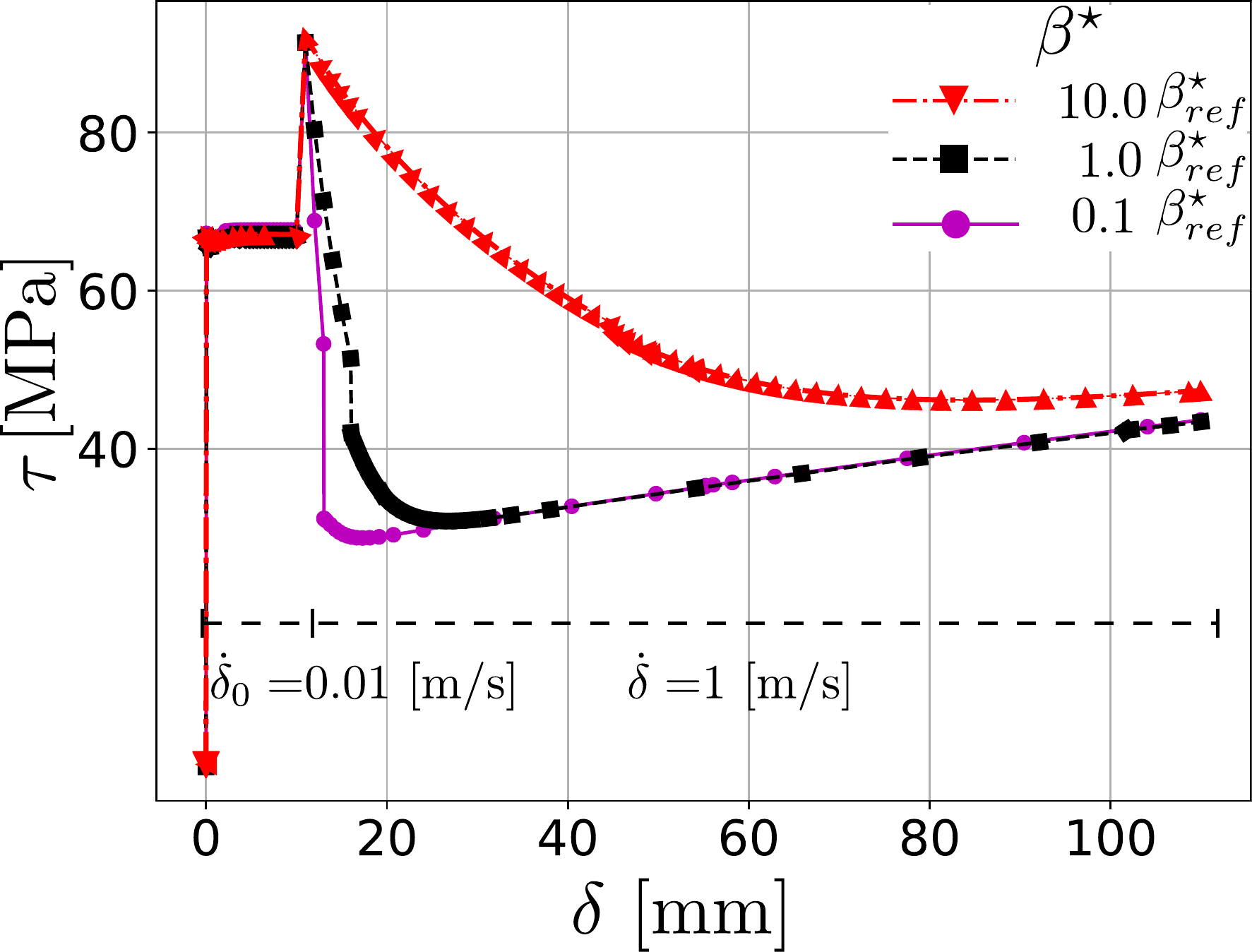}
  \end{minipage}%
  \caption{Left: Evolution of the fault's shear strength for a rapid change in shear rate $\dot{\delta}$ from 0.01 to 1.0 m/s, for different values of the size of the microstructure (Cosserat radius, $R$ in mm). The response is largely unaffected by the increase of the internal length. Right: Evolution of the fault's shear strength for a rapid change in shear rate $\dot{\delta}$ changes from 0.01 to 1.0 m/s, for different values of the mixture's compressibility parameter $\beta^\star$ (MPa$^{-1}$). Increasing mixture's compressibility leads to milder stress drops. }
  \label{ch: 5 fig: R_bstar_compare}
\end{figure}

\noindent Until now, a characteristic that is absent in the rate independent version of our model is the immediate positive frictional increase due to a sudden increase in the shearing velocity, due to viscous phenomena. These phenomena are considered in the frame of rate and state empirical models, through the ad hoc consideration of a velocity hardening term (see also \cite{Rice2001,Ruina1983,ruina1983slip,dieterich1992earthquake}, $\alpha>0$ for rate strengthening). 
Adopting rate and state friction as a reference, as shown in Figure \ref{ch: 5 fig: rate and state}, this means that our model misses some necessary physics at the microscale like a creep mechanism at the asperity scale level and a notion of a state variable $\psi$ describing the contact behavior over time. 
\\
\newline
\noindent 
It should be emphasized that the role of the state variable $\psi$ is ambivalent, in the sense that rate and state formulations do not necessarily respect the basic thermodynamic principles, since they mostly result from fittings to experimental results. Furthermore, we still lack data in the pressure and temperature ranges usually expected at the seismogenic depth, therefore, the parameters used in these laws are crucial concerning the prediction of the frictional evolution during a seismic event (see \cite{Rice2001,Rice2006}). Recently, emphasis has been given into the thermodynamically consistent derivation of granular material constitutive laws (see \cite{Alaei2021}). These novel constitutive material formulations, lead to thermodynamically consistent models with the introduction of granular inertia effects, that successfully capture the rate and state behavior observed in the laboratory experiments under the laboratory observed temperature range.\\
\newline
\noindent We include rate dependence in the present THM coupled model by introducing viscosity, which will lead to a strain rate hardening (or softening) description. In particular, we assume a Perzyna elasto-viscoplastic material introducing strain rate hardening effects through the use of a viscosity parameter $\eta$. The THM coupled model discussed here with the introduction of viscosity, can replicate the immediate effects of rate and state model, without the introduction of extra material parameters or the notion of an internal state variable indicating the contact history. In this section we investigate the role of viscosity in the frictional behavior of the fault during coseismic slip. A velocity stepping procedure is followed in which the fault gouge is initially slipping with a small seismic slip velocity ($\dot{\delta}_0=0.01$ m/s) for a 
small seismic slip displacement ($\delta=10$ mm). Then an immediate increase in the seismic slip rate is enforced in the model 
($\dot{\delta}=1.0$ m/s) to expose the rate dependence \cite[see also][]{dieterich1992earthquake,Rice2001,Ruina1983}. We continue shearing until the seismic slip $\dot{\delta}$ reaches a value of 100 mm. Then, we perform a series of parametric analyses to determine the influence of the viscosity parameter $\eta$, mixture compressibility 
$\beta^{*}$, the size of the microstructure (Cosserat radius $R$) and the seismic slip velocity $\dot{\delta}$, in the frictional response of the fault.\\
\begin{figure}[h]
  \centering
  \includegraphics[width=0.5\linewidth]{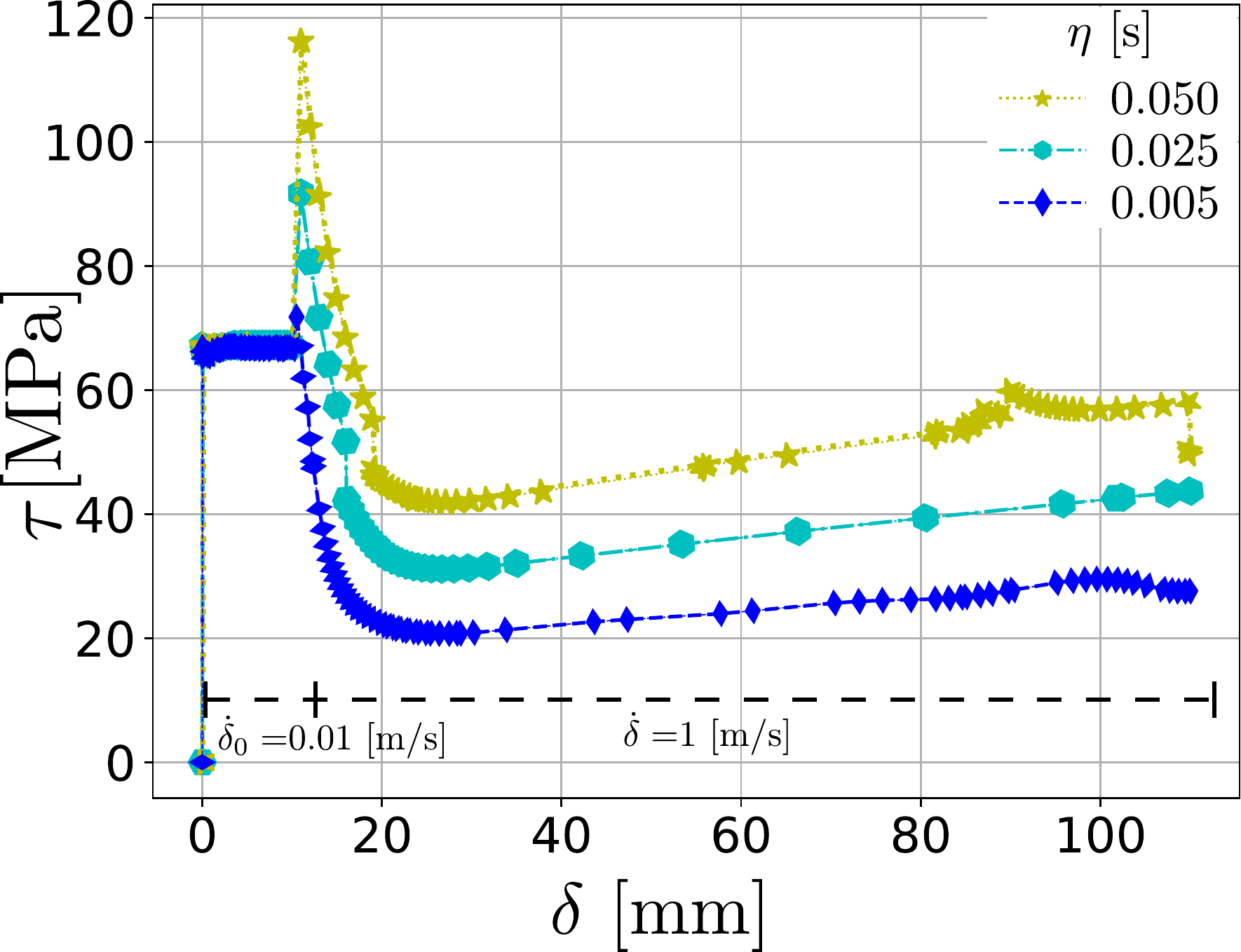}
  \caption{Evolution of the fault's shear strength for a rapid change in shear rate $\dot{\delta}$ changes from 0.01 to 1.0 m/s, for different values of the viscosity parameter $\eta$. For large values of the viscosity parameter $\eta$ (star curve), stress drops and kinks corresponding to stick slip events are presented.}
  \label{ch: 5 fig: eta_compare_1}
\end{figure}\\
\noindent In Figure \ref{ch: 5 fig: eta_compare_1}, we present the effect of the viscosity parameter in the frictional response of the fault for a sudden increase of the seismic slip rate from $\dot{\delta}_0=0.01$ m/s to $\dot{\delta}=1.0$ m/s. We observe that increasing the viscosity parameter $\eta$, the system becomes more sensitive to the sudden change of velocity reaching higher levels of peak frictional strength due to strain rate hardening. Three values are tested for the viscosity parameter $\eta=[0.005 s,0.025 s,0.05 s]$, while the other material parameters are taken from Table \ref{ch: 5 table:material_properties}. The lowest value corresponds to a meager increase in the fault's frictional strength due to the change in the shearing rate and is well in agreement with the results for the rate independent model. The frictional response initially reaches a minimum leading to velocity weakening. Afterwards, friction starts increasing due to pressure diffusion and strain-rate hardening. For the other values of the viscosity parameter the frictional strength increase is important. This change in the viscosity parameter also affects the minimum value of the frictional strength and its corresponding seismic slip displacement $\delta$.\\
\newline
\noindent The viscosity parameter $\eta$ introduced in the THM model accounts for the positive shear rate dependence coefficient ($\alpha$) of the rate and state model. Typical values for the non-dimensional coefficient $\alpha$ for faults, lie in the range of $10^{-4}-10^{-3}$, thus the lower values of $\eta$ of our analyses leading to the estimation of $\alpha = {\Delta \tau}/\left[\ln(\frac{\dot{\delta}}{\dot{\delta}_0})(\sigma_n-p_f) f\right]=2\;10^{-4}$ correspond well to the stress rate increase predicted by the rate and state model (assuming scaling of $\eta$ with $\dot{E}=\frac{\dot{\delta}}{h}=1000\;\text{s}^{-1}$).\\
\begin{figure}[h]
  \centering
  \includegraphics[width=0.5\linewidth]{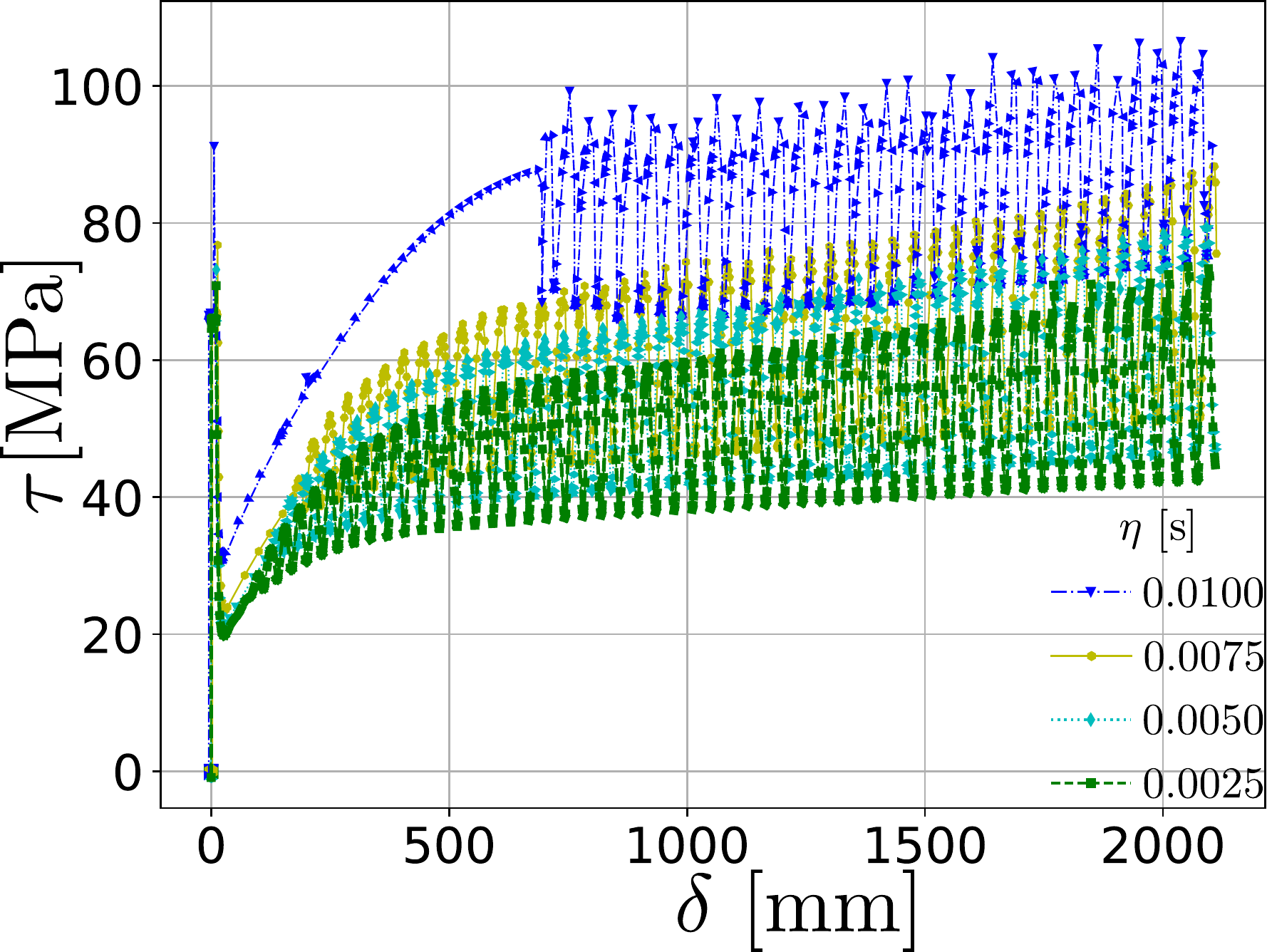}
  \caption{Evolution of the fault's shear strength for viscous parameter $\eta$ for common values of the rate and state parameter $\alpha=10^{-4}-10^{-3}$ and large seismic slip displacements $\delta=2$ m. The strain rate hardening leads to a constant increase of the center of the oscillations, that tend to reach the overstrength value. }
  \label{ch: 5 fig: eta_compare_2}
\end{figure}\\
\noindent Next, we apply a higher seismic slip displacement $\delta$ equal to 2 m during the stage of fast shear for the low values of the viscosity parameter, in order to capture the response of our model for larger seismic slips. In Figure \ref{ch: 5 fig: eta_compare_2}, we present the shear stress, seismic slip displacement $(\tau,\delta)$ evolution for shearing of a fault gouge with viscosity parameters $\eta$ in the range of [0.005 s - 0.010 s] for a seismic slip velocity, during the fast shear step, of $\dot{\delta}=1$ m/s. We note the small strain rate increase of the stress due to $\eta$ and the shear stress drop due to the apparent velocity weakening. The viscosity parameter enables the fault gouge to regain its overstrength in the latter analysis stages, since viscosity increases the localization width reducing frictional weakening due to thermal pressurization. We notice also that the oscillatory frictional response moves upward as seismic slip increases, while the oscillation maxima trace a curve of viscous evolution.\\
\newline
\noindent The oscillatory behavior of the frictional strength diagram could give rise to so called stick slip events in experiments. Thermal pressurization cannot completely halt strain-rate hardening, mainly due to the increased localization width the model exhibits, and the influence of the isothermal drained boundary conditions. We still can trace, however, a region of mild increases in the shear strength that can be used as an estimation of the characteristic weakening length, $D_c$, of the order of some centimeters (see also \cite{DiToro2011} among others). We note here that the estimation of the $D_c$ in the context of thermal pressurization is different than in the case of rate and state, in the sense that in the case of rate and state friction, $D_c$ is independent of the shearing velocity \cite[see][among others]{ruina1983slip}. In our case the characteristic distance depends on the shear velocity $\dot{\delta}$ the viscous parameter $\eta$, the size of the microstructure $R$ and the pressure and temperature diffusion lengths of the problem. \\
\newline
\noindent Next, we explore the influence of the size of the microstructure (Cosserat radius) $R$, in the evolution of the fault's frictional strength for a sudden change in the shearing rate from $\dot{\delta}=0.01$ m/s to 1 m/s (see Figure \ref{ch: 5 fig: R_bstar_compare}). We notice that an increase in the value of $R$ leads to higher shear overstrength due to the fast change in the shear rate. However, the post peak behavior changes little for values of $R$ varying from $0.1R_{ref}$ to $10R_{ref}$, where $R_{ref}$ is the Cosserat radius value in Table \ref{ch: 5 table:material_properties}. For $R=100 R_{ref}$ the increase in overstrength is substantial and the results show a faster regain of the strength and a higher minimum for the frictional strength after the initial apparent softening response.\\
\\
\noindent In Figure \ref{ch: 5 fig: R_bstar_compare} on the right, we present the influence of the mixture compressibility parameter $\beta^*$ in the frictional response of the fault for a value of the viscosity parameter $\eta=0.025$ s and a fast seismic slip velocity of $\dot{\delta}=1$ m/s. The parameter $\beta^*$ affects two terms in the equation \eqref{ch: 5 mass_balance_main}, namely the hydraulic diffusivity parameter $c_{hy}=\frac{\kappa}{\beta^*}$, where $\kappa$ is the permeability of the solid skeleton, and the term concerning the pressure decrease due to the porosity increase. Both these terms are affected the same by an increase (decrease) of $\beta^*$, however, the influence of the thermal pressurization term $\frac{\lambda^*}{\beta^*}$ decreases (increases) respectively. Taking the $\beta^*$ value for which we run the rate independent analyses as a reference value $\beta^*_{ref}$ (see Table \ref{ch: 5 table:material_properties}), this corresponds to a smoother (in the case of $\beta^*=10\beta^*_{ref}$), or steeper (in the case of $\beta^*=0.1\beta^*_{ref}$) decrease of the peak frictional strength during the initial stages of the slip. The parameter $\beta^*$ also controls the minimum frictional strength of the fault and the seismic slip $\delta$ for which, the frictional strength increase due to diffusion will become prevalent. The results agree qualitatively well with the behavior observed in \cite{Badt2020} for higher compressibilities due to the formation of gouge material at the initial stages (see next section).\\
\begin{figure}[h!]
  \centering
  \includegraphics[width=0.5\linewidth]{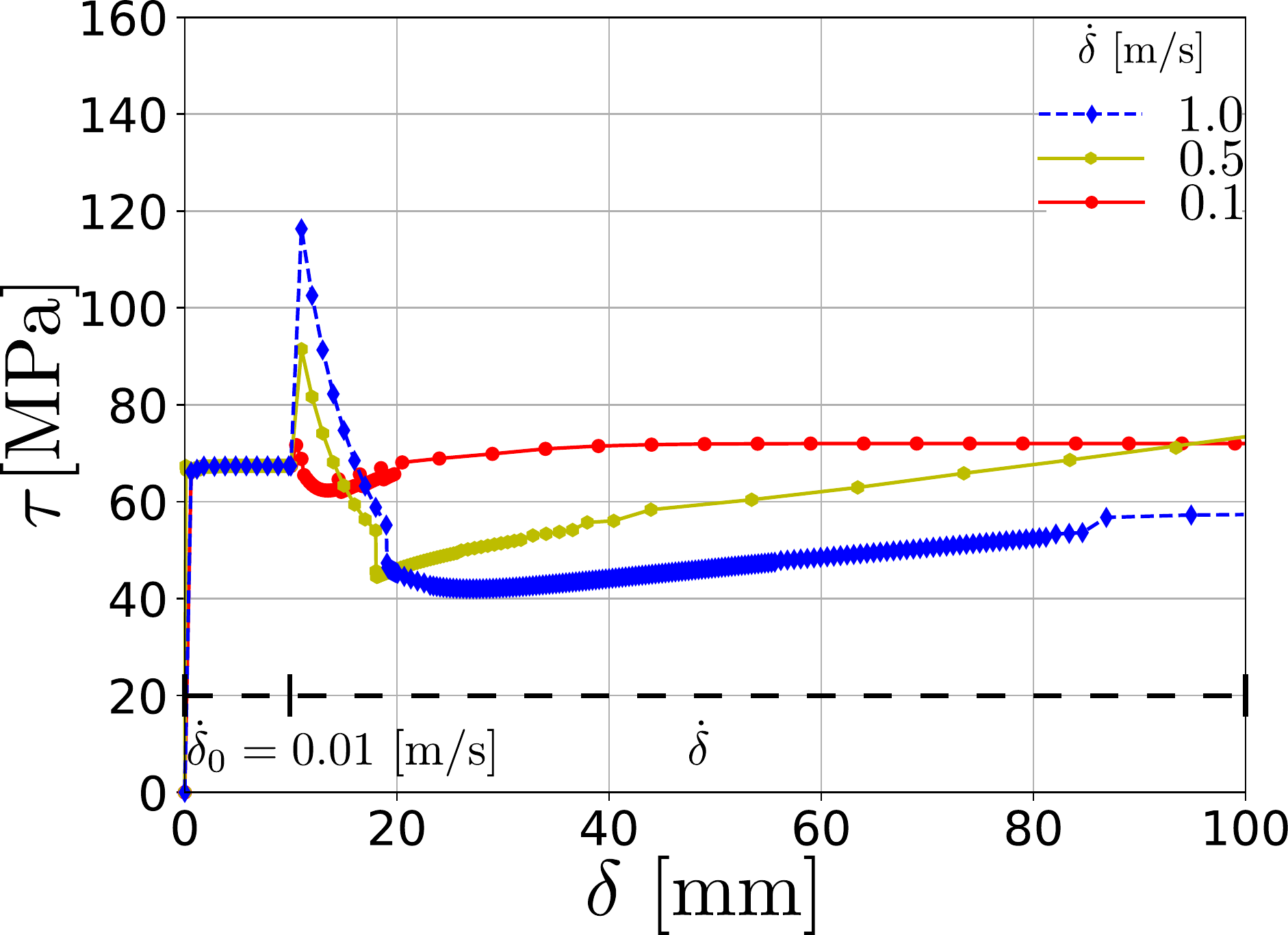}
  \caption{Frictional strength evolution of a Perzyna viscoplastic layer ($\eta=0.05$ s) under variable seismic slip velocities for a slip of $\delta=100$ mm.}
  \label{ch: 5 fig: vel_compare}
\end{figure}\\
\noindent Finally, in Figure \ref{ch: 5 fig: vel_compare}, we explore the influence of the shearing velocity in the frictional strength behavior of the fault for constant compressibility and viscosity parameters $\beta^s=\beta^s_{ref}$ and $\eta=0.05$ s respectively. After the initial slow shear $\dot{\delta}=0.01$ m/s, we vary the fast shear velocity from $\dot{\delta}$=0.1 m/s to $1.0$ m/s. The model exhibits two distinct behaviors, according to the prescribed seismic slip velocity values $\dot{\delta}$. For the low velocities $\dot{\delta}=0.1\sim 0.2$ m/s we observe periodic kinks in shear strength during the phase of shear strength increase. For higher velocities we observe a snap-back behavior during the apparent softening phase of the model. Furthermore a secondary stick slip event is observed during the strength regaining phase of the simulation due to the diffusion.\\
\newline
\noindent The above results suggest that with the introduction of viscosity, our physics-based model can describe the rate and state phenomenology. For different values of the viscosity parameter $\eta$ as well as the permeability $\beta^*$, stick slip behavior can be observed. This behavior suggests that our model of thermal pressurization together with a Perzyna viscoplastic law applied on a conceptually simple Drucker Prager yield criterion can capture a lot of the characteristics proposed by heuristic, phenomenological models like the rate and state friction law and its variations. Moreover, it can give further insights about the physical processes taking place during coseismic slip.
\section{Comparison with existing analytical solutions \label{ch: 5 Mase and Smith, Lachenbruch}}
\begin{figure}[h!]
  \centering 	 \includegraphics[width=0.5\linewidth]
  {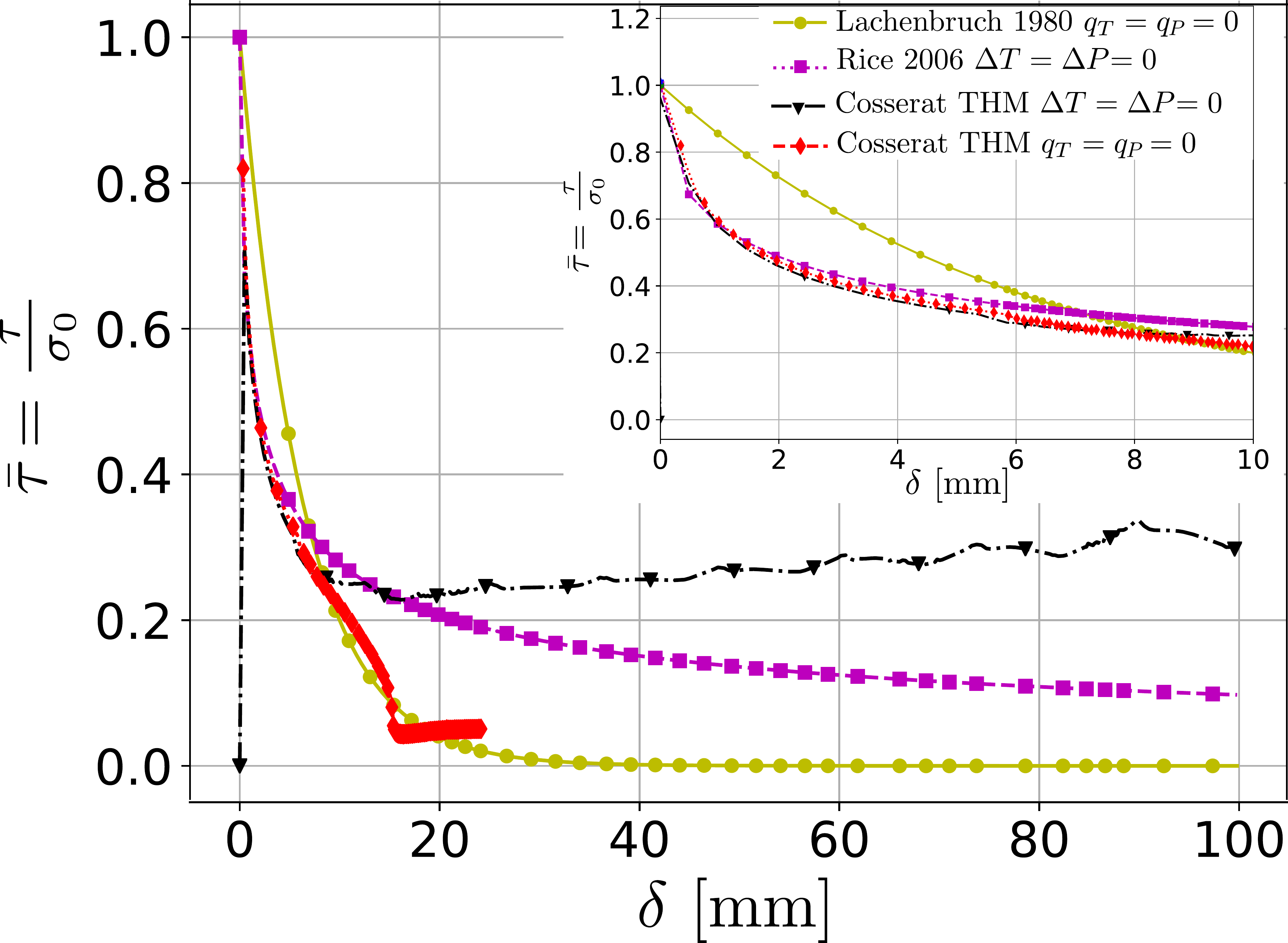}
  \caption{Comparison between the available numerical and analytical solutions for adiabatic undrained and isotropic drained conditions. The response of the Cosserat-THM model with isothermal drained boundary conditions (black-triangle line) lies close to the response of the slip on a plane solution provided in \cite{rice2006heating} (purple-square line), for small values of coseismic slip ($\delta\leq10$ mm). In the numerical model, localization is not constrained in a mathematical plane, leading to a steeper softening branch since more heat is produced in the yielding region enhancing thermal pressurization. Diffusion at the boundaries transfers heat and pressure away from the yielding region leading to partial strength regain causing a disagreement in the results. For the case of adiabatic undrained conditions, the Cosserat THM model (red-diamond line) eventually reaches the Lachenbruch solution (yellow-circle line).}
  \label{ch: 5 fig: tau_u_velocity_compare_1_ Solutions comparizon}
\end{figure}

\noindent We compare the nonlinear numerical solutions under adiabatic, undrained conditions ($q_T=q_p=0$) and isothermal, drained condtitions ($\Delta T=\Delta P=0$) to the reference analytical solutions obtained in bibliography for uniform shear of the layer (\cite{lachenbruch1980frictional}), as well as the concentrated shear on the mathematical plane (\cite{Mase1987,Rice2006}). The results are presented in Figure \ref{ch: 5 fig: tau_u_velocity_compare_1_ Solutions comparizon}.
\\
\newline
\noindent For adiabatic undrained conditions, the Cosserat numerical solution with THM couplings tends to the Lachenbruch solution after sufficiently large slip $\delta$. For isothermal drained conditions the numerical solution initially lies close to the solution for slip on a mathematical plane given in \cite{Rice2006}, however, the solution diverges as the seismic slip increases. This happens due to the diffusion at the isothermal drained boundary conditions, that restore part of the residual shear stress of the fault in later part of the analyses. 
The analytical solution for shear on a mathematical plane in \cite{Mase1987,Rice2006}, does not show this restrengthening behavior and the frictional oscillations. The reason is that the model in \cite{rice2006heating} assumes that slip localizes on a stationary mathematical plane inside an infinite (unbounded) layer. The model in \cite{Mase1987,Rice2006} neglects the ventilation phenomena that take place due to the interaction between the traveling PSZ (traveling thermal load) and the isothermal drained boundary conditions of the finite (bounded) fault gouge layer. In the companion paper (see \cite{Alexstathas2022b}), we will extend the solution of slip on a mathematical plane, for the bounded fault gouge under isothermal drained boundary conditions and a traveling PSZ. The extended model then, can represent better the main underlying physics captured by the detailed numerical simulations presented in this paper. \\
\newline
\noindent It is worth pointing out, that the traveling instability discussed here has been observed also in \cite{platt2014stability,rice2014stability}. However, the authors of that study applied periodic boundary conditions at the edges of the fault gouge instead of the isothermal drained conditions employed here. Furthermore, no connection was made in their model about the origin of these instabilities. Moreover, the authors did not comment on the effect of the traveling mode of strain localization (PSZ) on the layer's frictional response. In this paper we have identified the source of the traveling oscillations as an instability resulting from a Hopf bifurcation, and commented on them extensively. It is worth emphasizing, that while our analyses and the model proposed in \cite{Rice2006,platt2014stability,rice2014stability} agree well with the dynamic weakening role of thermal pressurization at the initial stages of the phenomenon, they diverge in later stages of the analyses in particular due to the role of the boundary conditions affecting thermal and hydraulic diffusion in the fault gouge. 
\section{ Comparison with existing experimental studies} \label{sec: ch: 5 experiments comparison }
\begin{figure}[h]
  \centering
  \includegraphics[width=0.9\linewidth]{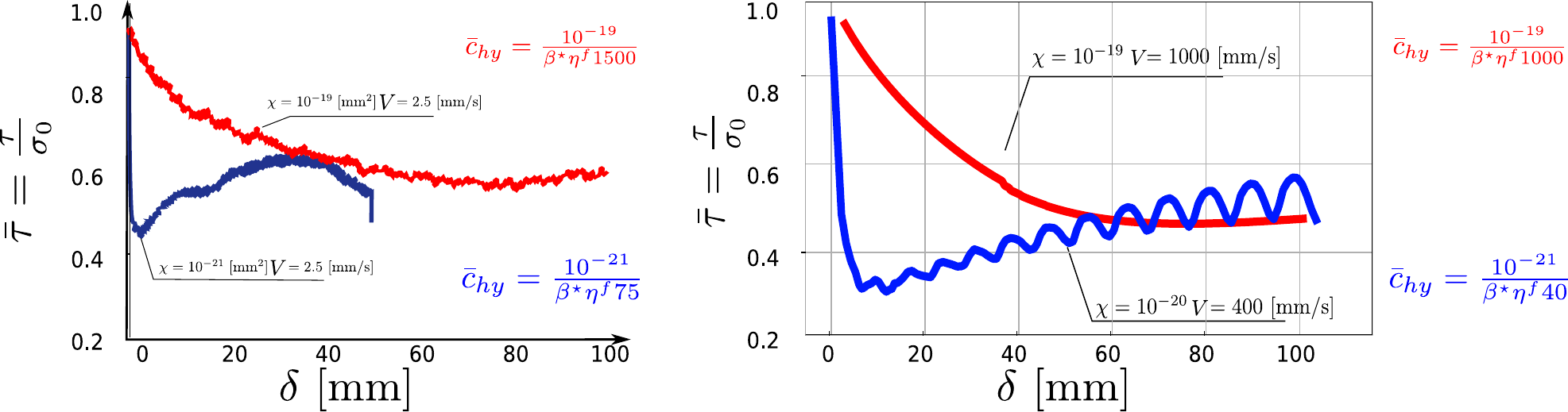}
  \caption{Qualitative comparison between experimental (left part of Figure) and numerical results (right part of Figure) examining the role of thermal pressurization. The experimental results were taken from \cite{Badt2020}. The curves correspond to low (blue color)  and high (red color) normalized hydraulic diffusivity values. In both cases shear velocity $\dot{\delta}$ and permeability $\chi$ influence heavily the results. Both the experiment and the numerical analysis predict frictional strength regain after the stress drop due to shear. The oscillatory frictional strength behavior of the numerical analyses is also present in the experiments of \cite{Badt2020} for low diffusivity values (blue curves). The experimental and numerical analyses are in qualitative agreement for normalized diffussivity values in the same parameter range.}
  \label{ch: 5 fig: Badt_compare}
\end{figure}

\noindent The results of the numerical analyses presented in section \ref{ch: 5 sec: Numerical results} are also observed in experiments (see \cite{Badt2020,DiToro2011,Rempe2020}). A great challenge seen in experimental studies about the role of thermal pressurization is the isolation of all other slip weakening mechanisms, in particular those of 
flash heating, silicate formation and thermal decomposition of minerals. Even bigger challenge is the replication of the exact ambient temperature and pressure conditions of the seismogenic zone. Moreover, the boundary conditions of the fault gouge, under which coseismic slip occurs are difficult to reproduce experimentally. Therefore, comparisons of analytical results to experimental findings are of a qualitative nature.\\
\newline
\noindent In \cite{DiToro2011} the authors, have accumulated a large body of experiments performed at rates and displacement ranges comparable to those during seismic slip. The experiments were performed in a range of normal stresses of the order of 0.6 to 20 MPa. The authors advocate that for seismic slip velocities of the order of $1$ m/s the frictional stress drop is around $0.2-0.4$ of the initial strength with higher drop as the normal stress increases. The strength drop is of the same magnitude in our analyses (see Figure \ref{ch: 5 fig: h_comparizon}). Furthermore, the obtained experimental frictional response presents oscillations, which could be attributed to the Portevin Le Chatelier phenomenon, due to a traveling strain localization during shearing of the specimen. The authors of this study also introduce the thermal weakening distance $D_{th}$ that scales with the applied effective pressure on the specimen. From extrapolation of the available data to the pressure ranges found in the seismogenic zone, they estimate the weakening distance $D_{th}$ to be of the order of centimeters. This observation is consistent with our analyses. We note here that the boundary conditions adiabatic, undrained vs isothermal, drained, the height of the specimen and the thermal and hydraulic diffusivities ($c_{th},c_{hy}$), affect the calculation of $D_{th}$. In particular, fully saturated (wet) specimens of larger height under isothermal drained conditions will drop to lower values of friction before the effect of the boundaries becomes noticeable (see also section \ref{ch: 5 Traveling_instability_new}). The ratio between thermal and hydraulic diffusivities (Lewis number $Le=\frac{c_{th}}{c_{hy}}$), controls the frictional weakening due to thermal pressurization. We note here that the value of $c_{hy}$ affects the characteristic time after which the effects of the boundaries will be felt in the frictional response. Therefore, the minimum value of the frictional response is controlled by the ratio of the diffussivities $c_{th},c_{hy}$ and the height of the layer.\\
\newline
\noindent In the experiments performed in \cite{Badt2020}, care was taken in order for the effect of thermal pressurization to be isolated from other weakening mechanisms. The authors of this study performed velocity stepping experiments in a rotary shear apparatus, with velocities of order $2.5\sim 5$ mm/s (well below the seismic range), under normal stresses of $20\sim 25$ MPa and confining pressure of $20\sim49$ MPa for final values of slip $\delta=2.6$ m. The specimen's height was $30$ mm. Their results suggest that frictional strength drops during the initial stages of thermal pressurization, while, depending on the evolution of the microstructure inside the fault (formation of fault gouge particles), partial regaining of the frictional strength is possible. The authors specify that steeper stress drop and restrengthening are observed in younger specimens that have not yet been subjected to large shear displacements ($\delta<1$ m) before the experiment. For specimens subjected to prior displacement the authors observe smoother stress drop and less tendency to regain frictional strength. They attribute this behavior to the formation of fault gouge inside the specimen that significantly affects the fault's compressibility coefficient $\beta^*$ and therefore, the hydraulic diffusivity $c_{th}$.\\
\newline
\noindent Comparing with our numerical results and extrapolating to the in situ pressure range (after rescaling), we observe that our model agrees very well qualitatively with the experimental findings for the younger specimens (see Figure \ref{ch: 5 fig: Badt_compare}). Since our model does not possess a memory mechanism to account for the damage of the microstructure and, therefore, for the change in permeability, it suffices to say that the older specimens could be modeled with higher values for the hydraulic diffusivity $c_{hy}$. An expansion to our model could be made by taking into account a microstructure evolution model as the one considered in \cite{collins2020cosserat}. We note here, that based on the scaled system \eqref{ch: 5 normalized_system_1}, the experimental results obtained during shearing of a specimen with height $\text{H}=30$ mm under a slip rate $\dot{\delta}=5$ mm/s, are comparable to our numerical experiments with fault gouge height $\text{H}=1$ mm and coseismic slip velocities in the range of $\dot{\delta}=100\sim300$ mm/s (see section \ref{ch: 5 Normalized_system_of_equations} for the appropriate scaling used). 
\section{Conclusions \label{Conlusions}}
\noindent In this paper we study the shearing of a fault gouge under large coseismic slip by considering a variety of seismic slip velocities $\dot{\delta}$ and seismic slip displacements $\delta$. First, we perform a linear stability analysis, indicating the possibility of traveling instabilities present in the medium. This result is further validated by the numerical results presented in section \ref{ch: 5 Traveling instability}. The effect of the seismic slip velocity $\dot{\delta}$ on the apparent softening of the layer is presented, as well as the influence of the different boundary conditions considering the energy and mass balance equation \eqref{ch: 5 Energy_balnca_eq}, \eqref{ch: 5 mass_balance_main} equations respectively. It is shown that at the initial stages of the shearing (seismic slip $\delta$=$10$ mm), after shear strain localization, frictional weakening occurs for the different cases of boundary conditions (see Figure \ref{ch: 5 fig: Tau_u_evolution_Loc_over_height_bcs_comparison}). The response, however, changes as the slip $\delta$ increases as shown in Figure \ref{ch: 5 fig: tau_u_velocity_compare_1_ Solutions comparizon}. Next, we investigate the influence of the seismic slip velocity $\dot{\delta}$ on the frictional behavior under isothermal drained conditions $(\Delta T=\Delta P=0)$ for a seismic slip of $100$ mm (Figure \ref{ch: 5 fig: tau_u_velocity_compare-fit}). It is shown that after the initial slip rate-dependent apparent softening, the layer tends to regain part of its strength $\tau$, as slip $\delta$ accumulates due to the existence of a limit cycle.\\
 \newline
 \noindent In the presence of isothermal drained boundary conditions we observe that the fault gouge regains part of its frictional strength. Frictional regain depends on the seismic slip rate $\dot{\delta}$, the thermal and hydraulic diffusivity properties $(c_{th},c_{hy})$ and the height $\text{H}$ of the gouge layer. Furthermore, we study the effects of the seismic slip rate on the evolution of the localization width $l_{loc}$. We notice the decrease of the localization width due to the increase in seismic slip velocity $\dot{\delta}$. This decrease indicates that the localization width is no longer dependent only on the Cosserat radius $R$, rather it is also dependent on the apparent softening due to thermal pressurization. There is an oscillatory behavior present in localization width particularly for low velocities. 
As the seismic slip rate increases, this oscillatory interaction reduces substantially. This further highlights the interplay between the different characteristic lengths in our model, in particular, between the Cosserat radius and the diffusion lengths.\\
\newline
\noindent Next, we study the evolution of the frictional resistance $\tau$ with seismic slip $\delta$ 
applying even larger seismic slip. In particular, we impose a realistic seismic slip displacement of $1$ m and seismic slip velocity of $1$ m/s. The oscillatory frictional behavior first described in Figure \ref{ch: 5 fig: tau_u_velocity_compare-fit} for smaller slip is also present in Figure \ref{ch: 5 fig: tau_u_velocity_compare_1} for large coseismic slip. These frictional oscillations are the result of the PSZ traveling inside the fault gouge. We notice that in Figure \ref{ch: 5 fig: tau_u_velocity_compare_1}, the frictional oscillations stabilize in frequency and amplitude. This implies a partial recovery of the frictional strength of the fault gouge, without it being explicitly implied by the mechanical behavior of the model as in the case of rate and state friction laws \cite[see][]{dieterich1992earthquake,Rice2001, Ruina1983}. Because of the traveling PSZ, only partial frictional regain is possible during coseismic slip. We explore further this behavior in Part II, see \cite{Alexstathas2022b}.\\
\newline
\noindent  During the isothermal drained analyses and the initial stages of the adiabatic undrained analyses, we observe  oscillations in the fault's frictional strength. This is a result of the traveling PSZ inside the fault gouge. The width and period of the oscillations depend on the height of the fault gouge. Frictional oscillations are the result of the high pressure and temperature diffusion gradients near the boundaries of the fault gouge. A similar result was achieved under periodic boundary conditions in \cite{rice2014stability}, however, the authors did not comment on the nature of the traveling instability (traveling PSZ). A very important consequence of the traveling shear strain rate instability (PSZ) is the fact that field observations regarding the fault gouge and its width may not be accurate. This is because it is impossible to know the strain rate history inside the fault gouge (see Figure \ref{ch: 5 fig: l_dot_gamma_final_1000-T_p_final_1000}). As a consequence the principal slip zone inside the fault gouge may be smaller than the one identified by current experimental methods. Traveling instabilities may also offer an explanation to the formation of parallel slip zones as the ones examined in \cite{nicchio2018development}.\\
\newline
\noindent\noindent We conclude our analyses, assuming a rate dependent material through the implementation of an elasto-viscoplastic Perzyna material with THM couplings. We notice that our model exhibits a strain rate hardening behavior after the sudden change of the shear rate. We further note that the value of the viscosity parameter $\eta$ controls important aspects of the simulation, such as the overstrength achieved due to the shearing rate, the existence and the magnitude of traveling instabilities (oscillations) in the solution, as well as the steady state the material reaches after sufficient shearing for varying shear rates. Overall, our Cosserat THM model with viscoplasticity is exhibiting a lot of the characteristics of a rate and state phenomenological model, without using an internal frictional state $\psi$ variable and its evolution law. The procedure of thermal pressurization has still a lot of potential explaining the frictional strength drop during an earthquake, together with the fault nucleation.
\\
\newline
\noindent Finally, it can be shown that the results of the above numerical analyses agree qualitatively well with the recent experimental results obtained in \cite{Badt2020}, where the thermal pressurization mechanism was studied in isolation to different frictional weakening mechanisms. Considering the evolution of the fault's frictional strength with accumulated seismic slip, we observe that a single value for the characterization of the critical slip distance $D_c$ is not possible. Namely, according to Figure \ref{ch: 5 fig: tau_u_velocity_compare_1}, the fault strength drops almost immediately after a slip of few centimeters to a minimum value close to 1/3 of the initial strength. This result agrees with extrapolations from experimental data to higher confining stresses (see \cite{DiToro2011}). However, the final value around which the residual strength oscillates, is reached for a slip distance of $0.4$ m (see Figure \ref{ch: 5 fig: tau_u_velocity_compare_1}). The latter result agrees with the estimation of \cite{Rempel2006,Rice2006}.\\
\newline    
\noindent Our results are in contrast to the classical models of THM proposed in \cite{lachenbruch1980frictional,rice2006heating} that assume only monotonic reduction of the shear frictional strength during the seismic event. In order to understand the extreme difference between the predictions of the model described in \cite{lachenbruch1980frictional,Rice2006} and our numerical results, we revisit in Part II \cite[see][]{Alexstathas2022b}, the main assumptions of the theory of \cite{Rice2006} considering a traveling mode of strain localization and by modifying the boundary conditions.
 We will examine how each one of these parameters affect the theoretical predictions and will further justify our numerical results with Cosserat theory and THM couplings. 
\section*{Acknowledgments}
\noindent The authors would like to acknowledge the support of the European Research Council (ERC) under
the European Union’s Horizon 2020 research and innovation program (Grant agreement no. 757848
CoQuake).
\let\clearpage\relax
\appendix
\begin{appendices}
\section{Constitutive relations\label{Appendix D}}
\noindent The devolopement of the thermo-elasto-plastic constitutive ralations that follow is based on \cite{sulem2011stability} and \cite{Rattez2018a}. Since we follow a small strain approach, the strain rate and the curvature rate tensor can be decomposed into their elastic, plastic and thermal parts. Large displacements are then taken into account through an updated Lagrangian approach. In what follows we make the assumption that the curvature tensor stays unaffected by a change of temperature. Therefore strain rate and curvature rate tensors are decomposed as in \cite{lemaitre2020mecanique}:
\begin{align}
\dot{\gamma}_{ij} =& \dot{\gamma}^e_{ij}+\dot{\gamma}^p_{ij}+\dot{\gamma}^{th}_{ij}, \nonumber\\
\dot{\kappa}_{ij} =& \dot{\kappa}^e_{ij}+\dot{\kappa}^p_{ij}
\label{app D: eq: final_functional_equation}
\end{align}
Thermal strain rates can be expressed as $\dot{\gamma}^{th}_{ij}=\alpha \dot{T}\delta_{ij}$, where $\alpha$ is the thermal expansion coefficient. For the calculation of the plastic strain rate, we first define a yield function $F=F(\tau_{ij},\sigma_{ij},\gamma^p,\epsilon^p_v)$, which we assume to be dependent only on the first and second stress tensor invariants as well as the deviatoric and spherical parts of the accumulated plastic strain tensor $F=F(\tau,\sigma,\gamma^p,\epsilon^p_v)$. A more complete approach in a thermodynamical framework that takes into account grain breakage and the consequent evolution of the internal lengths can be found in \cite{collins2020cosserat}.  Following standard arguments of elasto-plasticity and by use of the consistency condition $\dot{F}$ we obtain:
\begin{align}
&\tau_{ij} = C^e_{ijkl}\left(\dot{\gamma}_{ij}- \dot{\gamma}^p_{ij}- \dot{\gamma}^{th}_{ij}\right),
\label{app D: tau_ij_rate_relation}\\
&\mu_{ij} = M^e_{ijkl}\left(\dot{\kappa}_{ij}- \dot{\kappa}^p_{ij}\right),
\label{app D: mu_ij_rate_relation}\\
&\dot{\gamma}^p_{ij}=\dot{\gamma}^p\frac{\partial Q}{\partial \tau_{ij}},
\label{app D: gamma_pl_ij}\\
&\dot{\kappa}^p_{ij}=\dot{\gamma}^p\frac{\partial Q}{\partial \mu_{ij}},
\label{app D: kappa_pl_ij}\\
&\dot{F}=\frac{\partial F}{\partial \tau_{ij}}\dot{\tau}_{ij}+\frac{\partial F}{\partial \mu_{ij}}\dot{\mu}_{ij}+\frac{\partial F}{\partial \gamma^p}\dot{\gamma}^p+\frac{\partial F}{\partial \epsilon^p_v}\dot{\epsilon}^p_v=0,\\
&\dot{F}=\frac{\partial F}{\partial \tau}\dot{\tau}+\frac{\partial F}{\partial \sigma}\dot{\sigma}+\frac{\partial F}{\partial \gamma^p}\dot{\gamma}^p+\frac{\partial F}{\partial \epsilon^p_v}\dot{\epsilon}^p_v=0.
\label{app D: General_yield_function}
\end{align}
\noindent Where by $Q,\dot{\gamma}^p$ we denote the plastic potential and the plastic multiplier respectively. We note that in the present context a common criterion for both Cosserat stresses and moments has been assigned to the material. We continue by defining the hardening modulus $H_s$ as:
\begin{align}
H_s=&-\frac{\partial F}{\partial \gamma^p}.
\end{align}
Assuming a linear dependence of the yield and plastic potential functions to $\tau,\sigma$ as is the case in a Drucker-Prager material, which we will later use in the numerical analyses, the following relations hold for the plastic multiplier and the rate of volumetric plastic strain:
\begin{align}
\dot{\gamma}^p=\dot{\gamma}^p \;\text{ and }\;
\dot{\epsilon}^p_v = \beta \dot{\gamma}^p
\end{align} 
\noindent where $\beta$ is the dilatancy angle.
Multiplying \eqref{app D: tau_ij_rate_relation} by $\frac{\partial F}{\partial \tau_{ij}}$ and \eqref{app D: mu_ij_rate_relation} by $\frac{\partial F}{\partial \mu_{ij}}$ then adding together and taking advantage of the fact that $\frac{\partial F}{\partial \tau_{ij}}\dot{\tau}_{ij}+\frac{\partial F}{\partial \mu_{ij}}\dot{\mu}_{ij} = \frac{\partial F}{\partial \tau}\dot{\tau}+\frac{\partial F}{\partial \sigma}\dot{\sigma}$, the consistency condition yields:
\begin{align}
&\dot{\gamma}^p=\frac{<1>}{H_p}\left(\frac{\partial F}{\partial \tau_{ij}}C^e_{ijkl}(\dot{\gamma}_{kl}-\alpha\dot{T}\delta_{kl})\right)+\frac{\partial F}{\partial \mu_{ij}}M^e_{ijkl}\dot{\kappa}_{kl}.\\
\intertext{Simplifying the notation we get:}
&\dot{\gamma}^p = \frac{<1>}{H_p}(b^F_{kl}(\dot{\gamma}_{kl}-\alpha\dot{T}\delta_{kl})+b^F_{kl}\dot{\kappa}_{kl}),
\label{app D: lambda_dot}\\
\intertext{with}
&H_p = \frac{\partial F}{\partial \tau_{ij}}C^e_{ijkl}\frac{\partial Q}{\partial \tau_{kl}}+\frac{\partial F}{\partial \mu_{ij}}M^e_{ijkl}\frac{\partial Q}{\partial \mu_{kl}}+H_s\\
<1> =& \begin{cases}
      1 & \text{if $F=0$ and $\dot{\gamma}^p>0$}\\
      0 & \text{otherwise}
    \end{cases}
\intertext{and}
&b^F_{kl} =\frac{\partial F}{\partial \tau_{ij}C^e_{ijkl}}, \\
&b^Q_{ij} = C^e_{ijkl}\frac{\partial Q}{\partial \tau_{kl}},\\
&b^FM_{kl} = \frac{\partial F}{\partial \mu_{ij}M^e_{ijkl}}, \\
&b^QM_{ij} = M^e_{ijkl}\frac{\partial Q}{\partial \mu_{kl}}.\\
\end{align}
Using \eqref{app D: gamma_pl_ij},\eqref{app D: kappa_pl_ij} and \eqref{app D: lambda_dot} in \eqref{app D: tau_ij_rate_relation} we obtain:
\begin{align}
\label{Constitutive_relation_thermo_elasto_plastic_final}
&\dot{\tau}_{ij}=C^{ep}_{ijkl}\dot{\gamma}_{kl}+D^{ep}_{ijkl}\dot{\kappa}_{kl}+E^{ep}_{ijkl}\dot{T}\delta_{kl}\\
&\dot{\mu}_{ij}=M^{ep}_{ijkl}\dot{\kappa}_{kl}+L^{ep}_{ijkl}\dot{\gamma}_{kl}+N^{ep}_{ijkl}\dot{T}\delta_{kl}\\
\intertext{with}
&C^{ep}_{ijkl} =C^e_{ijkl}-\frac{<1>}{H_p}b^Q_{ij}b^F_{kl},\nonumber\\
&D^{ep}_{ijkl} =-\frac{<1>}{H_p}b^Q_{ij}b^FM_{kl},\nonumber\\
&E^{ep}_{ijkl} =-\left(C^e_{ijkl}-\frac{<1>}{H_p}b^Q_{ij}b^F_{kl}\right),\nonumber\\
&L^{ep}_{ijkl} =-\frac{<1>}{H_p}b^QM_{ij}b^F_{kl},\nonumber\\
&M^{ep}_{ijkl} = \left(M^e_{ijkl}-\frac{<1>}{H_p}b^QM_{ij}b^FM_{kl}\right),\nonumber\\
&N^{ep}_{ijkl} =\frac{<1>}{H_p}b^QM_{ij}b^F_{kl}.
\end{align}
\end{appendices}
\nocite{*} 
\typeout{}

\bibliography{/home/alexandrosstathas/Documents/PhD_topics/PAPER_II/PAPER_II_Submission_docs_Arxiv/biblio/bibliography_clean.bib}

\end{document}